\documentclass[fleqn,usenatbib]{mnras}

\usepackage{newtxtext,newtxmath}

\usepackage[T1]{fontenc}

\usepackage{graphicx}	
\usepackage{amsmath}	

\usepackage{comment}

\newcommand{\fnl}{f_{\rm NL}}
\newcommand{\hfnl}{\hat{f}_{\rm NL}}
\newcommand{\hfnlh}{\hat{f}_{\rm NL}^{100}}
\newcommand{\hfnlz}{\hfnl^{0}}

\newcommand{\gpng}{\textsc{goliat-png}}
\newcommand{\Mpch}{h^{-1}{\rm Mpc}}
\newcommand{\Gpch}{ h^{-1}{\rm Gpc} }

\title[Fixed, Paired \& Matched ICs for PNG]{
\begin{minipage}{7.03 in}
\vskip -0.4 in
\begin{flushright}
{\rm \small IFT-UAM/CSIC-22-47}
\end{flushright}
\end{minipage}\\
\vskip 0.0in
\vspace{0.1 cm} 
Validating galaxy clustering models with  Fixed \& Paired and Matched-ICs simulations: application to Primordial Non-Gaussianities}

\author[S. Avila \& A. G. Adame]{
Santiago Avila$^{1,2,3}$\thanks{E-mail: santiagoavilaperez@gmail.com} and 
Adrian Gutierrez Adame$^{1,2}$\thanks{adrian.gutierrez@uam.es}
\\
$^{1}$ Departamento de F\'isica Te\'orica, Universidad Aut\'onoma de Madrid, 28049 Madrid (Spain) \\
$^{2}$ Instituto de F\'isica Teorica UAM-CSIC, c/ Nicolás Cabrera 13-15, , 28049 Madrid \\
$^{3}$ Institut de Fisica d’Altes Energies (IFAE), The Barcelona Institute of Science and Technology, Campus UAB, 08193 Bellaterra (Barcelona) Spain
}

\date{Accepted 5 Dec 2022. Received Nov 2022; in original form April 2022}

\pubyear{2023}

\begin{document}
\label{firstpage}
\pagerange{\pageref{firstpage}--\pageref{lastpage}}
\maketitle

\begin{abstract}
The {\it Fix} and {\it Pair} techniques were designed to generate simulations with reduced variance in the 2-point statistics by modifying the Initial Conditions (ICs). In this paper we show that this technique is also valid when the initial conditions have local Primordial non-Gaussianities (PNG), parametrised by $\fnl$, without biasing the 2-point statistics but reducing significantly their variance. We show how to quantitatively use these techniques to test the accuracy of galaxy/halo clustering models down to a much reduced uncertainty and we apply them to test the standard model for halo clustering in the presence of PNG. Additionally, we show that by {\it Matching} the stochastic part of the ICs for two different cosmologies (Gaussian and non-Gaussian) we obtain a large correlation between the (2-point) statistics that can explicitly be used to further reduce the uncertainty of the model testing. For our reference analysis ($\fnl=100$, $V=1 [\Gpch]^3$, $n= 2.5\times 10^{-4}[\Mpch]^{-3}$, $b=2.32$), we obtain an uncertainty of $\sigma(\fnl)=60$ with a standard simulation, whereas using {\it Fixed} [{\it Fixed}-{\it Paired}] initial conditions it reduces to $\sigma(\fnl)=12$ [$\sigma(\fnl)=12$]. When also {\it Matching} the ICs we obtain $\sigma (\fnl)=18$ for the standard case, and $\sigma (\fnl)=8$ [$\sigma (\fnl)=7$] for {\it Fixed} [{\it Fixed}-{\it Paired}]. The combination of the {\it Fix}, {\it Pair} and {\it Match} techniques can be used in the context of PNG to create simulations with an effective volume incremented by a factor $\sim 70$ at given computational resources.
\end{abstract}

\begin{keywords}
cosmology: large-scale structure of the Universe -- galaxies: haloes  -- cosmology: theory -- cosmology: inflation
\end{keywords}



\section{Introduction}
\label{sec:intro}

Cosmological simulations are routinely used in cosmology, specially in studies of the Large-Scale Structure (for a recent review, see \citealt{Angulo_review}). They can be used to help developing analysis methodology \citep[e.g.][]{Ross17,Chan22,Avila22}, validate and compare the performance of different tools \citep[e.g.][]{Avila20,Alam21,Rossi21}, understand the impact of observational systematic effects \citep[e.g.][]{Carnero22,Spinelli22} or to compute covariance matrices \citep[e.g.][]{Manera13,Avila18,Zhao21}. 

When testing models of galaxy clustering, we are limited by the intrinsic uncertainty associated to the simulation we are using, given by cosmic variance and shot-noise. In order to reduce this, one could increase arbitrarily the simulated volume. However, given that we typically want to focus on the halo mass-range given by a reference galaxy survey, if we keep the mass resolution constant, this process rapidly becomes prohibitively computationally expensive. 

In order to alleviate this, the \textit{Fix \& Pair} technique was proposed in \citet{Angulo16}. \textit{Fixing} consists on setting the amplitude of the overdensity modes in the Initial Conditions (ICs) to its expectation value (without any variance), whereas {\it Pairing} consists on running a second simulation with the initial phases {\it Inverted} (shifted by $\pi$) with respect to the first one. 
This technique has extensively studied and was shown to recover unbiased statistics for late-time observables such as the 1-point (halo mass function, luminosity functions, etc.), 2-point (power spectrum and correlation function of dark matter and biased tracers such as halos), 3-point statistics, etc. Additionally, many of those statistic show a decrement of their variance, which is particularly drastic in the 2-point statistics \citep[][]{Angulo16,FVN18,Unitsim,FnP_cov,Maion22}. This reduction of the variance, which is particularly concentrated at large scales in the power spectrum, can be used to validate models of the Large-Scale Structure (LSS) with increased accuracy for the same computational cost.

Another way to reduce the effective variance when validating models with simulations is to {\it Match} the stochastic part of the ICs in simulations with different cosmologies. This induces correlation between the measured clustering, with the cosmic variance partly cancelling when, for example, dividing the clustering of two simulations. This is mentioned in several works like \citet{halofit_ext}, however, the correlation coefficient was not explicitly applied to reduce the uncertainty in the analysis. 
The correlation between {\it Matched}-ICs of simulations has been explicitly used in the CARPool method \citep[][]{CarPool,DESIsim2} but, in that case, between high-fidelity mocks and approximate mocks, in order to retrieve unbiased precise clustering statistics from the approximate mocks. Here, we propose a framework to explicitly use the correlation between simulations with {\it Matched}-ICs across different cosmologies to reduce the expected variance of the summary statistic and to constrain models to an increased accuracy. 

Both of those techniques are particularly promising for large-scale clustering. One interesting observable at very large scales is the local Primordial Non-Gaussianities (PNG), parametrised by $\fnl$, which will serve us to illustrate the methods proposed here. 

PNG is one of the observables of cosmic inflation: whereas the simplest models predict a level of local-PNG below $\mathcal{O}(\fnl)=1$, more complex models (in particular multi-field inflation), predict larger PNG \citep{Creminelli04,Pajer13,Byrnes}. Whereas Planck CMB experiment \footnote{\url{https://www.cosmos.esa.int/web/planck}} constrained local-PNG down to $\fnl=-0.9\pm5.1$ \citep{PlanckPNG}, the precision is near the expected cosmic variance limit, with some room for improvement with polarisation experiments \citep{CMBpol}. However, future surveys of the LSS (or their combination), such us SKAO\footnote{\url{https://www.skatelescope.org/}} \citep{SKA_cosmo}, LSST-Rubin\footnote{\url{https://www.lsst.org/}} \citep{LSST_science_09,LSST_science18} or SphereX\footnote{\url{https://spherex.caltech.edu/}} \citep{SphereX_cosmo,SphereX_cosmo_cross} are expected to break the $\sigma(\fnl)=1$ barrier \citep{Yamauchi14,Alvarez,dePutter}, opening a new window to understand inflation. 

One of the most promising ways to constrain PNG is by using the scale-dependent bias that it induces at very large-scales on the power spectrum \citep[][]{Dalal08,Slosar08,Matarrese}. 
Over the last decade, measurements of the 2-point functions of the LSS has led to increasingly precise constrains \citep{Slosar08, 2012Ross, PhysRevD.89.023511, 2015Ho, PhysRevLett.113.221301, 2019Castorina}, with the most precise measurements given by the clustering of eBOSS\footnote{\url{https://www.sdss.org/surveys/eboss/}} quasar clustering: $\fnl=-12\pm21$ \citep{Mueller22,Rezaie21}. Recently, \citet{Cabass_2022} put similar constraints ($\fnl=-33\pm28$) by combining power spectrum and bispectrum using BOSS \footnote{\url{http://www.sdss3.org/surveys/boss.php}} galaxies, opening another promising path for more voluminous surveys.

By {\it Fixing} the amplitude of the initial conditions, we are actually introducing a type of non-Gaussianity, since we are substituting the Rayleigh distribution of the amplitude of the modes in {\it Normal} Gaussian simulations by a Dirac delta. Hence, the first goal of this paper is to demonstrate that the {\it Fix} (\& {\it Pair}, although this part does not pose any {\it a priori} issue) technique can be safely used with local Primordial Non-Gaussianities. 

The second goal is to demonstrate with a practical example how to validate models of halo clustering (extendable to galaxy clustering) using {\it Fixed \& Paired} simulations to a reduced uncertainty. 
Finally, we also show how to explicitly use the correlation of the clustering of simulations with different cosmologies but {\it Matched}-ICs to further reduce the uncertainty associated.
We demonstrate the potential of these techniques by testing the PNG halo power spectrum model derived from the peak-background split model together with a universal halo mass function \citep{Dalal08,Slosar08}. Alternatively, we also show that the same techniques can be applied to significantly reduce the uncertainty on measured nuisance parameters associated to PNG halo bias ($p$ or $b_\phi$, see \autoref{sec:model}). In similar lines, the recent work of \citet{Zennaro21} \& \citet{Maion22} also showed how to put tight priors on Gaussian ($\fnl=0$) Lagrangian bias parameters when using {\it Fixed} and {\it Paired} simulations. 

For this study, we developed the \gpng\ simulation suite: a series of simulations with {\it Normal}, {\it Inverted}, {\it Fixed} and {\it Fixed-Inverted} initial conditions for 41 different ICs realisations and for $\fnl=0$ \& $\fnl=100$, which we will make publicly available.

The paper is organised as follows. In \autoref{sec:ICs}, we describe how the initial conditions of our simulations are generated, discussing: the standard Gaussian case, the implementation of local PNG and the implementation of the {\it Fix}, {\it Pair} and {\it Match} techniques. In \autoref{sec:halo_clustering}, we describe the \gpng\ simulation suite, built for this paper, and study their mean halo power spectrum $P(k)$ and variance $\sigma(k)^2$. Then, \autoref{sec:bias} is dedicated to the (scale-dependent) halo bias, describing the modelling and fitting procedure applied to the \gpng\ simulations. In \autoref{sec:results}, we show the main results of the paper: how the {\it Fix}, {\it Pair} and {\it Match} techniques can be used to validate the PNG halo clustering model down to a reduced uncertainty. We summarise our findings in \autoref{sec:summary} and discuss future prospects in \autoref{sec:outlook}.

\section{Fixed \& Paired PNG Initial Conditions}
\label{sec:ICs}

The main ingredients of the methodology discussed in the paper stems from choosing the way we generate the ICs that we input to the $N$-Body gravity solver to generate the simulations. Here we discuss how to generate the ICs in a normal Gaussian case, how this gets modified with the Primordial Non-Gaussianities, how to set {\it Fixed} and {\it Paired} ICs and the role of the stochastic part of the ICs in the {\it Matching}.

\subsection{Gaussian Initial Conditions}
\label{sec:ICs_normal}

In the standard simulations, the initial conditions are generated from Gaussian realisations of the overdensity field $\delta(k)$ whose variance is given by the power spectrum P(k). 
Equivalently, one can express those initial conditions in terms of the primordial gravitational potential $\Phi$. For clarity when introducing later the PNG and the {\it Fix} technique, we will express the ICs in terms of the latter. 

The late time linear overdensity and the primordial gravitational potential can be related by:
\begin{equation}
    \delta(k,z) = \alpha(k,z) \Phi(k) \, ,
    \label{eq:poisson}
\end{equation}
where we have defined
\begin{equation}\label{eq:alpha}
    \alpha(k,z)=\frac{2D(z)}{3\Omega_{m}(z)}\frac{c^{2}}{H_{0}^{2}}\frac{g(0)}{g(z_{\rm rad})}k^{2}T(k) \, ,
\end{equation}
where $D(z)$ is the growth factor normalized to $D(z=0) = 1$ and $T(k)$ the transfer function (with $T(k\to 0 )=1$, LSS convention). The factor $\frac{g(z_{\rm rad})}{g(0)}=1.31$ (for the cosmology described in \autoref{sec:sims}) takes into account the difference between the LSS normalization of $D(z)$ with respect the early-time normalisation to $D(z) =1/(1+z)$ during matter-domination.

Then, ICs can be generated as random realisations of a Gaussian distribution:
\begin{equation}
    \Phi_i(k) \curvearrowleft \mathcal{N_{\mathbb{C}}}(\Phi_i(k);\mu=0, \sigma^2=P(k) / \alpha^2)\, ,
\end{equation}
with $\mathbb{C}$ reminding us its complex nature in Fourier space and where we have removed the explicit dependence of $\alpha$ on $z$ and $k$ for simplicity. We use the symbol '$\curvearrowleft$' to denote random variables (left) that are sampled from a probability distribution function (right). Alternatively, $\Phi$ can be decomposed into modulus and phases $\lvert \Phi\lvert$, $\varphi$. 
The modulus, then, follows a Rayleigh distribution: 
\begin{equation}
    \lvert \Phi (k)\lvert_i \curvearrowleft  \mathcal{P}_{\rm Rayleigh} (\lvert \Phi (k)\lvert) = \frac{\lvert \Phi (k)\lvert \alpha^2}{P(k)} \exp \left( \frac{-\lvert \Phi (k)\lvert^2 \alpha^2}{2\ P(k) }\right)\, , 
    \label{eq:rayleigh}
\end{equation}
and the phases $\varphi$, a uniform distribution between $0$ and $2\pi$
\begin{equation}
    \varphi_i \curvearrowleft \frac{1}{2\pi} \Theta_{[0,2\pi]}(\varphi)\, .
    \label{eq:phases}
\end{equation}

The $i$ denotes that it is a random realisation, given by a seed in our code. This seed is changed for different realisations of the simulations. If we {\it Match} the random seed for different cosmologies ($\fnl=0$ and $\fnl=100$ in our case), we will be using the same phases $\varphi_i$ and the same relative excess/decrement in amplitude of modes with respect to their expectation value: $\lvert \Phi (k)\lvert_i \cdot \alpha/\sqrt{P(k)\cdot \pi/2}$. Hence, sharing great part of the noise realisation. 

Finally, these perturbations of the Newtonian potential $\Phi$ (or equivalently in the
overdensity field) are used to compute the displacements and velocities of the
dark matter particles through the second-order Lagrangian Perturbation Theory (2LPT, \citealt{2LPT}). For that, we use the public code \textsc{2LPTic} \citep{2LPTic_paper,2LPTic_code} to generate the initial conditions.

\subsection{Local Primordial Non-Gaussianities}

For a Gaussian random field, all the information is contained in the 2-point statistics, with the higher order ones vanishing. 
For most models of inflation, the primordial field $\Phi$ follows a nearly Gaussian distribution. 
Then, small Primordial Non-Gaussianities are parametrised by $\fnl$, which quantifies the level of non-Gaussianity associated to the primordial bispectrum. Other higher contributions are quantified by other parameters such as $g_{\rm NL}$ or $\tau_{\rm NL}$, which we do not explore here.

The primordial bispectrum can take different shapes, depending on the ratios of the different modes considered.
The simplest configuration to study is the local PNG, which modifies the primordial potential perturbations as \citep{localPNG,Salopek91}:

\begin{equation}
    \Phi\left(\textbf{x}\right) = \Phi_G\left(\textbf{x}\right) + \fnl^{\rm loc} \big( \Phi_G\left(\textbf{x}\right)^2 - \langle \Phi_G ^2 \rangle \big) ,
    \label{eq:quadratic}
\end{equation}
where $\Phi_G$ represents a Gaussian gravitational field, as described in the previous subsection, and $\Phi$ now represents the Non-Gaussian gravitational field, which can be related to the non-Gaussian linear overdensity by \autoref{eq:poisson}. We will drop the `$^{\rm loc}$' label for most cases in this paper, as we will be referring by default to local PNG. We remark that \autoref{eq:quadratic} describes a modification in configuration space, whereas most of the equations described earlier refer to Fourier space. 
It has been shown \citep{localPNG} that this type of non-Gaussianities leads to a primordial bispectrum given by 

\begin{equation}
    \mathcal{ B}(k_1,k_2,k_3) = \fnl \cdot 2\left( \mathcal{P}(k_1)\mathcal{P}(k_2) + \mathcal{P}(k_2)\mathcal{P}(k_3) +\mathcal{P}(k_3)\mathcal{P}(k_1) \right) . 
\end{equation}

The local PNG (and other types of PNG) are already implemented in \textsc{2LPTic} as described in \citet[][]{2LPT-PNG}.

\subsection{Fix and Pair techniques}

\citet{Angulo16} proposed to {\it Fix} the amplitude of the initial perturbations to their expectation value in the ICs in order to reduce the variance associated to simulations. Mathematically, this means substituting the Rayleigh PDF (\autoref{eq:rayleigh}) by a Dirac delta:
\begin{equation}
    \lvert \Phi(k) \lvert_i \curvearrowleft \frac{1}{2\pi}\delta_D({\lvert \Phi(k) \lvert} -  \sqrt{P(k) \cdot \pi/2}  /\alpha  )\, .
	\label{eq:dirac}
\end{equation}

From now on, we will refer to these simulations as {\it Fixed}, whereas the simulations whose ICs follow a Rayleigh distribution will be tagged as {\it Normal}. We note that, whereas there is not any stochasticity left in the amplitude of the ICs modes, the phases $\varphi_i$ will still be random realisations (\autoref{eq:phases}). 

By  describing the ICs in terms of $\Phi$, \autoref{eq:dirac} can be directly applied to the Gaussian and local-PNG cases. In the local-PNG case we apply \autoref{eq:dirac} to the Gaussian component and subsequently we apply the quadratic correction in \autoref{eq:quadratic}. 

Both \autoref{eq:quadratic} and \autoref{eq:dirac} represent a modification to the Gaussian ICs. Hence, it is one of the goals of this paper to demonstrate that these two techniques can be safely used together and that the {\it Fixed} technique does not introduce any significant spurious PNG signal. On \autoref{sec:halo_clustering}, we will validate the use of {\it Fixed} local-PNG simulations in the context of halo clustering and in particular on the measurement of scale-dependent bias. This is the most promising avenue to detect local-PNG with the LSS and its study is central to this paper. We leave a more detailed validation of the simulations to the Appendix \ref{app:validation}, where we compare the dark matter and halo power spectra and bispectra for both {\it Fixed} and {\it Normal} simulations with and without PNG.

\citet{Angulo16} also proposed to {\it Pair} simulations by running one {\it Original} simulation together with another identical {\it Inverted} simulation in which we simply shift the phases $\varphi$ by $\pi$ radians: 

\begin{equation}
    \varphi_{\it Inverted} = \varphi_{\it Original} + \pi\, .
\end{equation}

Where we recall that the phases are cyclic with a period of $2\pi$.
\citet{Angulo16}
showed that averaging the statistics from these two {\it Paired} simulations can cancel out further terms in the variance of the {\it Fixed} simulations of dark matter 2-point statistics. A more recent and detailed study on the way {\it Fix} and {\it Pair} reduces the variance also of biased tracers is presented in \citet{Maion22}.

The {\it Fix} and {\it Pair} techniques are not implemented in the public version of \textsc{2LPTic}, but we included by modifying a few lines both in the Gaussian and Non-Gaussian versions of the code. 

\section{The halo clustering with Fixed \& Paired PNG simulations}
\label{sec:halo_clustering}
\subsection{The Goliat-PNG simulation suite}
\label{sec:sims}

\begin{table} 
  \centering
  \caption{The \gpng\ cosmological and setup parameters. The cosmological parameters \citep{wmap7}:  $\Omega_{\rm m}$, $\Omega_{\rm b}$ and $\Omega_\Lambda$ are the average densities of total matter, baryonic matter and vacuum energy in units of the critical density today, $H_0$ is the Hubble parameter, $\sigma_8$ is the 
  rms of the matter fluctuations at $8~h^{-1}{\rm Mpc}$ and $n_{\rm s}$ is the spectral index of the primordial power spectrum. Simulation parameters: redshift of the snapshot used in this paper, volume of the simulation box, number of particles in the simulation and particle mass resolution. 
  }
 \label{tab:cosmo}
\begin{tabular}{|l|r|}
\hline
   $\Omega_{\rm m}$             &    0.27 \\
   $\Omega_{\rm b}$              & 0.044 \\
   $\Omega_\Lambda$             &    0.73 \\
   $h \equiv H_0$/(100 km\,s$^{-1}$\,Mpc$^{-1})$ &  0.7\\
   $\sigma_8$                   &    0.8 \\
   $n_{\rm s}$                  &    0.96 \\
   \hline
   $z_{\rm snap}$  & 1.0  \\
   Volume       &   $(1000 h^{-1}{\rm Mpc})^3$  \\
   $N$             &    512$^3$ \\
   $m_{\rm p}$    &    $5.58\times10^{11}h^{-1}M_{\odot}$ \\
\hline
	\end{tabular}
\end{table}

The \gpng\ simulation suite builds up on previous existing runs used for \citet{halogen} and \citet{Wang20}.
For this reason, we take for reference the cosmology of WMAP-7 \citep{wmap7} shown \autoref{tab:cosmo} and a resolution of $512^3$ particles in a $V=1 [\Gpch]^3$ box. 
We run the initial conditions with the \textsc{2LPTic}\footnote{\url{cosmo.nyu.edu/roman/2LPT} \citep{2LPTic_paper,2LPT-PNG,2LPTic_code}} (see also \autoref{sec:ICs}). Then, we evolve those ICs with the public gravity solver \textsc{Gadget2} \footnote{\url{https://wwwmpa.mpa-garching.mpg.de/gadget/} \citep{Gadget2}} down to $z=1$. Finally, we identify halos with the Amiga Halo Finder \footnote{\url{http://popia.ft.uam.es/AHF/} \citep{AHF}} and use halos with a minimum of $10$ particles, which gives us a number density of $n=2.5\cdot 10^{-4} {\rm Mpc}^{-3}h^3$.

We focus on the redshift $z=1$ snapshot, as running the simulations down to $z=1$ requires about half of the time of that required to run them to $z=0$. Additionally, large observational surveys with promising forecasts on PNG will focus on the $z\ge 1$ Universe in order to probe larger volumes. We note that, even though the used cosmology is ruled out by current observational constraints \citep{Planck,PlanckPNG} and the limited resolution could have an effect on derived parameters such as $p$ (PNG response, see \autoref{sec:model}), the focus of this work is on the methodology to increase the the precision of the derived constraints by using the {\it Fix}, {\it Pair} \& {\it Match} techniques, and to verify that the {\it Fixing} can be applied to PNG simulations in the same way as we do for the {\it Normal} ones.

We run 41 different realisations for the $\fnl=0$ and $\fnl=100$ cosmologies for 4 cases:

\begin{itemize}
    \item {\it Normal}-{\it Original}.
    \item {\it Normal}-{\it Inverted}.
    \item {\it Fixed}-{\it Original}.
    \item {\it Fixed}-{\it Inverted}.
\end{itemize}

\subsection{Mean halo power spectrum with the Goliat-PNG simulations}
\label{sec:Pk}

\begin{figure*}
	\includegraphics[width=\columnwidth]{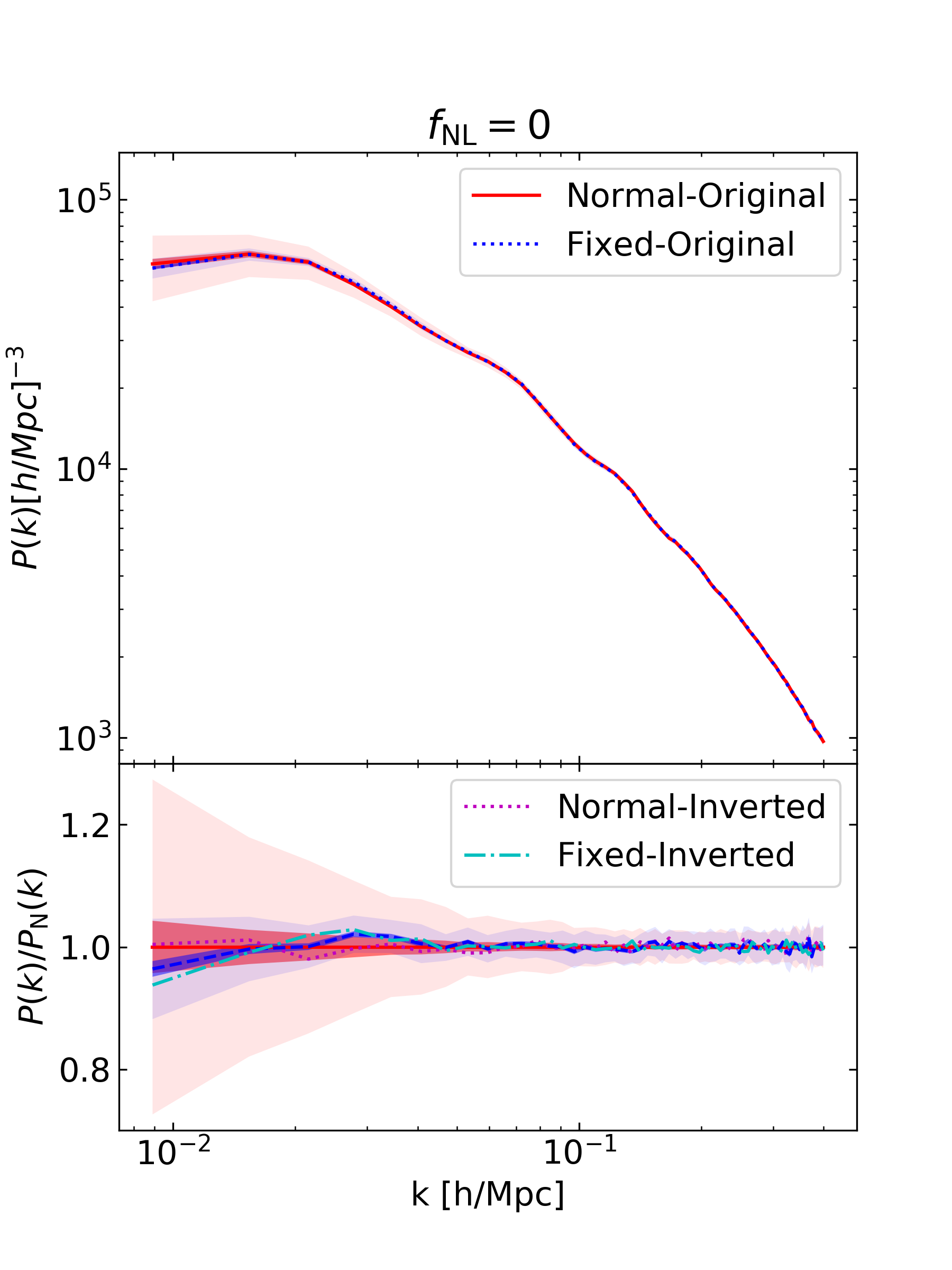}\includegraphics[width=\columnwidth]{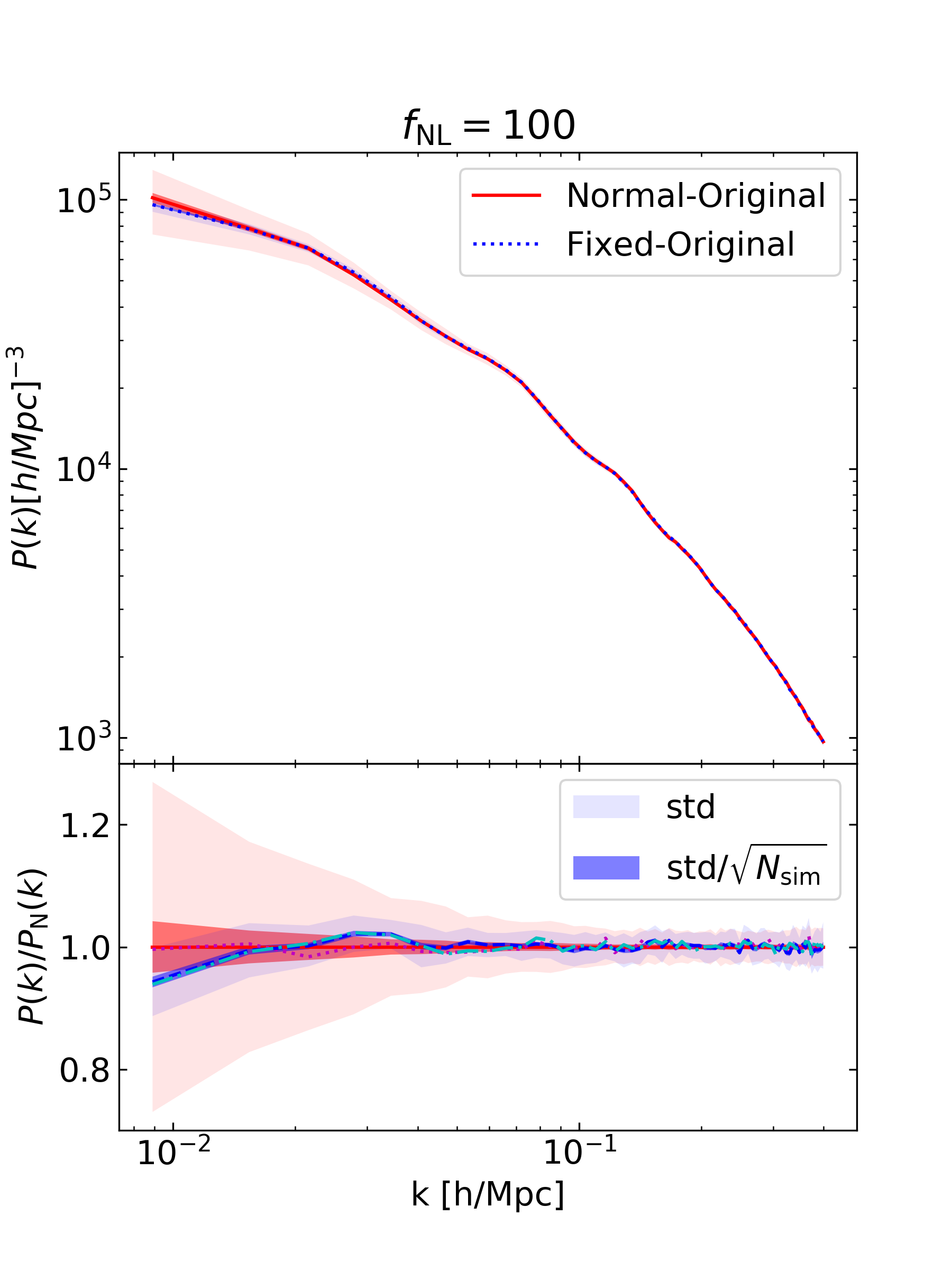}
    \caption{Mean halo power spectrum of the \gpng\ $N$-body simulations. On the left we show the results for the $\fnl=0$ case, whereas on he right we show the results for $\fnl=100$. We compare the {\it Normal} simulations  to the {\it Fixed} simulations from the 41 {\it Original} ICs (red and blue, respectively), whereas in magenta and cyan we also show their respective {\it Inverted} realisations (only at the bottom panels). The light-shaded regions represent the standard deviation over the $N_{\rm sim}=41$ realisations (our reference error for the rest of the paper), whereas the dark-shaded regions represent the estimated uncertainty on the ensemble average. We find that the ratios of the {\it Fixed} simulations to the {\it Normal} simulations are nearly identical for the $\fnl=0$ and $\fnl=100$ cases, and compatible with unity. Hence, we conclude that the {\it Fix} \& {\it Pair} technique can also be safely used for PNG simulations. 
    }
    \label{fig:Pk}
\end{figure*}

Here, we will validate the usage of {\it Fixed} PNG simulations in order to study the halo power spectrum. We leave a more detailed validation, considering other statistics, to Appendix \ref{app:validation}.

We compute the power spectrum of the \gpng\ halo catalogues with \textsc{Nbodykit}\footnote{ \url{https://github.com/bccp/nbodykit}, \citep{nbodykit}}. 
In \autoref{fig:Pk}, we represent the halo power spectrum of the \gpng\ simulations for $\fnl=0$ and $\fnl=100$. We find that, for both cases, the average power spectrum is consistent between the {\it Normal}  simulations and the {\it Fixed} simulations. Whereas this has already been extensively studied for the case of $\fnl=0$ \citep{Unitsim,FVN18,Maion22}, 
this is the first time this is shown for the local PNG case with $\fnl\neq0$. This is important since the {\it Fixed} technique itself introduces a type of non-Gaussianity in the initial conditions. Nevertheless, \autoref{fig:Pk} shows that the ratios of the {\it Normal} to {\it Fixed} results (within the noise level) are found  very similar for the $\fnl=0$ and $\fnl=100$ cases. Hence, we find no evidence that the {\it Fix} technique for local PNG needs any special consideration. Additionally, it is worth noting that the goal of the paper focuses on the power of a single realisation to constrain cosmological parameters. Hence, our reference scatter is the light-shaded region.

The {\it Normal}-{\it Original} and the {\it Normal}-{\it Inverted} simulations are statistically equivalent by construction, hence, they have the same expectation values for any statistics. It is only when combining both that we expect modifications on the variance of, for example, the power spectrum. The same happens with the  {\it Fixed}-{\it Original} and the {\it Fixed}-{\it Inverted} cases. To verify this, we also show in the bottom panels of \autoref{fig:Pk} the comparison between the {\it Original} and the {\it Inverted} simulations, finding negligible statistical differences. For the rest of the paper, we will always consider the average of all the {\it Normal} ({\it Original}+{\it Inverted}) or {\it Fixed} ({\it Original}+{\it Inverted}) simulations, when referring to the average statistics. The next subsection will describe how we deal with the variance.

\subsection{Halo power spectrum variance of the Goliat-PNG simulations}
\label{sec:variance}
\begin{figure*}
	\includegraphics[width=\columnwidth]{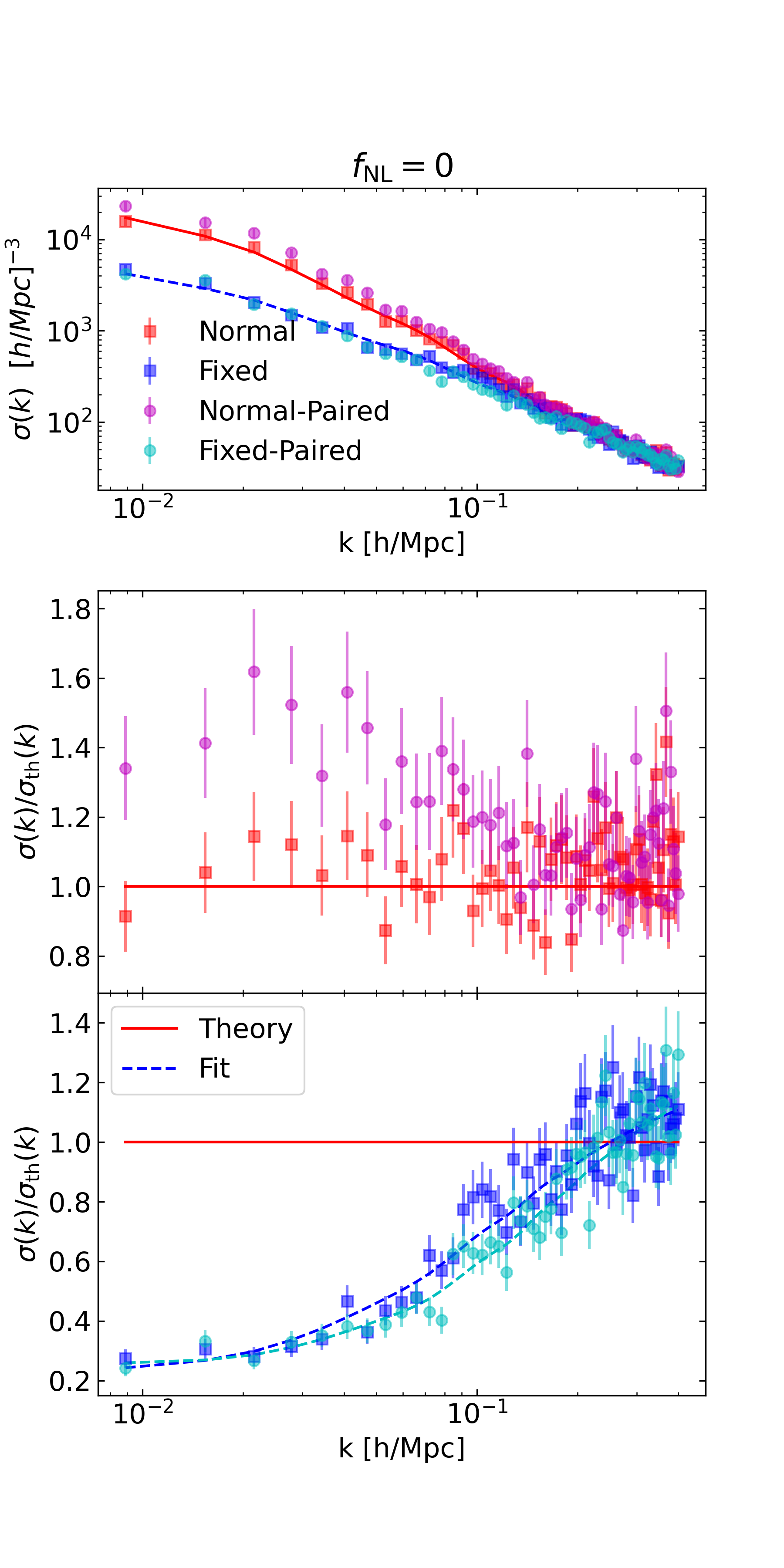}\includegraphics[width=\columnwidth]{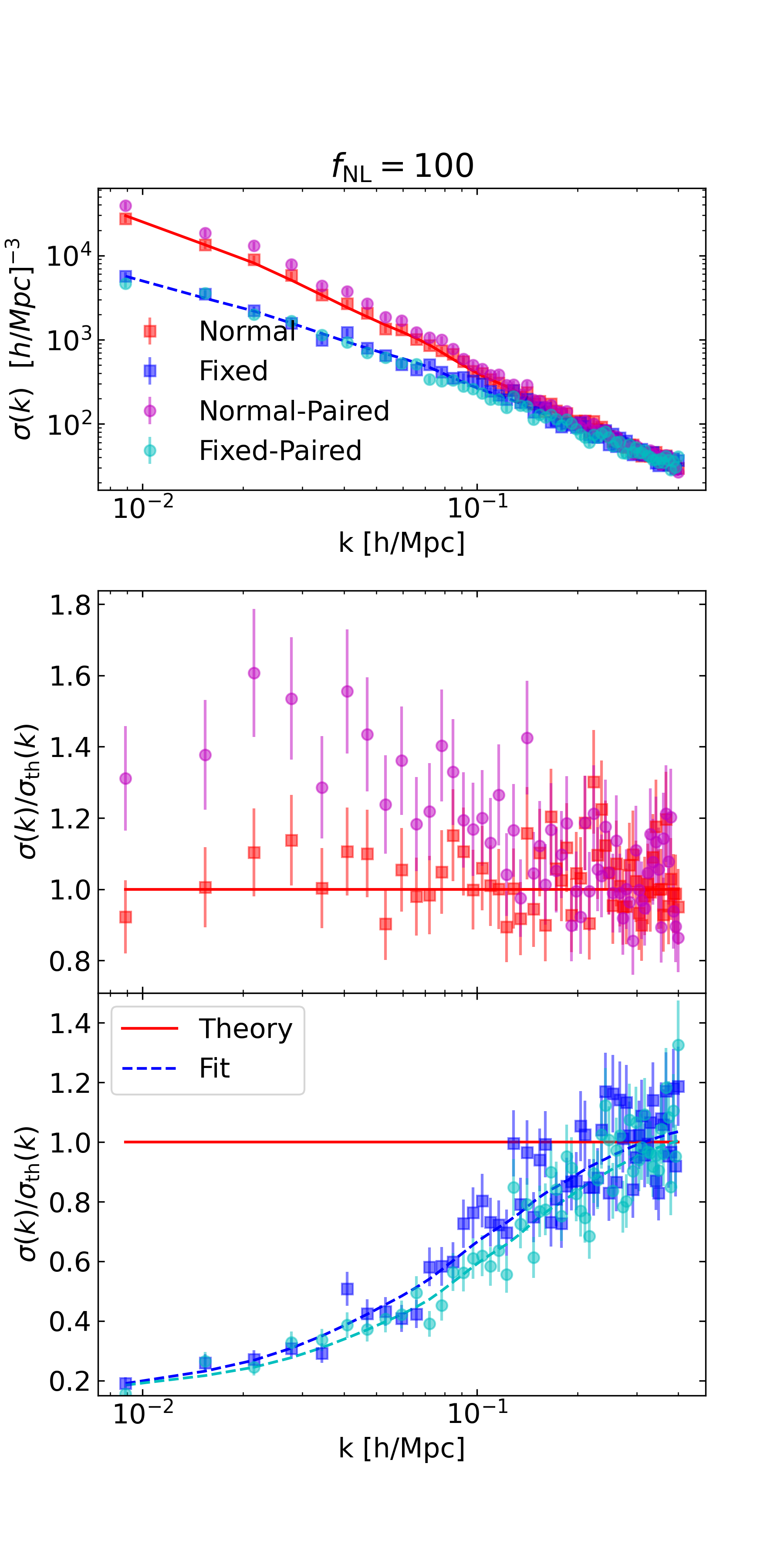}
    \caption{Standard deviation of the halo power spectrum measured in the \gpng\ (points with error bars given by \autoref{eq:errerr}) compared to the theoretical prediction for {\it Normal} simulations (red solid line, \autoref{eq:sigma_th}) and proposed phenomenological fits for the {\it Fixed} and {\it Fixed}-{\it Paired} simulations (dashed lines, \autoref{eq:sigma_fix}, \autoref{tab:fits}). Left: $\fnl=0$. Right: $\fnl=100$. Top: absolute $\sigma$. Bottom: Ratio with respect to the theoretical prediction for the {\it Normal} simulations.}
    \label{fig:sigma}
\end{figure*}

For the case of a Gaussian density field with a given power spectrum $P(k)$ and discretised with a Poisson-law to a number density $n$, the expected variance over a volume $V$ is given by \citep{FKP,Blake19}

\begin{equation}
    \sigma_{\rm th}(k)^2 = \frac{4\pi^2}{V\ k^2\ \Delta k} \Bigg( P(k)  + \frac{1}{n} \Bigg)^2 \, \, .
    \label{eq:sigma_th}
\end{equation}
Where $\Delta k$ is the bin width over which the power spectrum $P(k)$ has been measured. 

The {\it Fixed} and {\it Paired} technique was introduced in \citet{Angulo16} precisely to reduce that variance on simulations. On the one hand, fixing the amplitude of the primordial overdensities, reduces de variance of the late-time overdensities. On the other hand, pairing two simulations with inverse phases cancels out certain terms, reducing again the variance \citep[see][]{Angulo16,Maion22}.

From this point, when we refer to results derived from the P(k) of {\bf {\it Paired}} simulations (either {\it Normal} or {\it Fixed}), it means that we have taken the average of each simulation with its {\it Inverted} partner:

\begin{equation}\label{eq:pair}
    P_{\rm Paired, i}(k) = \frac{1}{2}(P_{\varphi_i}(k)+P_{\varphi_i+\pi}(k))\, .
\end{equation}

When we compute the standard deviation of {\it Paired} simulations, we first average over the couple, then compute the standard deviation and finally, multiply by $\sqrt{2}$ to compensate for doubling the volume\footnote{However, in \autoref{sec:results}, the {\it Fixed}-{\it Paired} results will use $\sigma_{\rm FP}/\sqrt{2}$ as we will compare results to the scatter of the pairs, which have this reduction intrinsically.}: 

\begin{equation}
    \sigma_{\rm Paired} = \sqrt{2}\cdot {\rm std}\Bigg( \frac{1}{2} \big(P_{\varphi_i}(k)+P_{\varphi_i+\pi}(k)\big)\Bigg)\, .
    \label{eq:sigma_paired}
\end{equation}

In \autoref{fig:sigma}, we show the standard deviation obtained from the 41 cases of {\it Normal}, {\it Fixed}, {\it Normal}-{\it Paired} and {\it Fixed-Paired} simulations (the last two, using \autoref{eq:sigma_paired}). On the solid red line, we also represent the theoretical Gaussian expectation, given by \autoref{eq:sigma_th}.

We find that the scatter of the {\it Normal} simulations (red) is well described by \autoref{eq:sigma_th}, at least at large scales. On the other hand, the {\it Fixed} simulations present a large reduction at small $k$, as expected and previously found. Remarkably, whereas the {\it Fixed-Paired} case shows a slight further reduction at intermediate scales over the {\it Fixed} case, the {\it Normal-Paired} case shows an increment of scatter with respect to the {\it Normal} case. 

When comparing the $\fnl=0$ and $\fnl=100$ simulations, we find very similar results in the ratios of the simulation variances to the expected Gaussian variance (\autoref{eq:sigma_th}). Hence, by combining the results from \autoref{fig:Pk} \& \autoref{fig:sigma}, we conclude that the {\it Fix} \& {\it Pair} technique can be use for local Primordial Non-Gaussianities in the same manner it is used for the Gaussian simulations. 
 
In \citet{FnP_cov} they studied in detail the covariance matrices for halo clustering of {\it Fix} \& {\it Pair} simulations. They find that, for the power spectrum, the covariance matrix remains diagonal as it does in the standard case. They also show that the variance reduction cannot be only accounted for by a constant reduction in the cosmic variance term and that the exact shape of this reduction depends on redshift, halo mass, etc. Very recently, \citet[][]{Maion22} developed a framework to  predict from perturbation theory, the scatter reduction expected for a given sample. Hence, we also expect in our case (with PNG), that the noise reduction will depend on the specific sample. 

In this study, we fit the scale-dependent variance reduction with the following  phenomenological expression:
 
\begin{equation}
    \sigma_{\rm r}(k) = \sqrt{\frac{4\pi^2}{V\ k^2\ \Delta k}} \Bigg( P(k)\cdot \Big[ R_{\rm cv} - (1-R_{\rm cv})\cdot \frac{2}{\pi}{\rm arctan}\Big(\frac{k}{k_{\rm soft}}\Big)\Big] + \frac{1}{n}\cdot f_{\rm sn} \Bigg)\, ,
    \label{eq:sigma_fix}
\end{equation}
with the aim of having a less noisy version of the uncertainties presented in \autoref{fig:sigma}. Note that this is simply a slight modification of \autoref{eq:sigma_th}, where we introduce a smooth step function (with softening scale $k_{\rm soft}$) between a large-scale reduced cosmic variance (by a factor $R_{\rm cv}$) and the standard value at small scales (large-$k$). We also consider a super-Poisonian shot-noise factor $f_{\rm sn}$. This last factor is used to account for an excess scatter observed at large $k$, which is not specific to the {\it Fixed} simulations as it is also found in the {\it Normal} simulations. The existence of excess shot-noise due to halo formation being a non-Poissonian process is well documented in the literature \citep{Hamaus10,Baldauf13,Desjacques}. 
We note, however, that this affects only large $k$, whereas we will focus later on the small-$k$ ($k<0.09 h^{-1}$Mpc). Nonetheless, we fit the expression above for all the scales $k<0.4\Mpch$, finding good agreement. 

The fits with \autoref{eq:sigma_fix} to the {\it Fixed} and {\it Fixed}-{\it Paired} variances is shown in \autoref{fig:sigma} (blue for {\it Fixed}, cyan for {\it Fixed}-{\it Paired}) and the best-fit values of the parameters are presented in \autoref{tab:fits}. 
The error on the scatter ($\Delta(\sigma)$) plotted in \autoref{fig:sigma} and considered for the fits has been estimate as \citep{LehmCase98}:

\begin{equation}
    \Delta(\sigma) = \frac{1}{\sqrt{2 (N_{\rm sim}-1)}} \sigma\, .
    \label{eq:errerr}
\end{equation}

\begin{table}
	\centering
	\caption{Best fit parameters of \autoref{eq:sigma_fix} for the standard deviation measured on the {\it Fixed} simulations and the {\it Fixed} \& {\it Paired} (using \autoref{eq:sigma_paired}) simulations. These parameters and equation will be used as the error for most of the analysis of the paper.}
	\label{tab:fits}
	\begin{tabular}{lccc} 
		\hline
		 & $R_{\rm cv}$  & $k_{\rm soft}$ [$h/$Mpc] & $f_{\rm sn}$\\
		\hline
		$\fnl=0$, $\sigma_{\rm F}$ & 0.141 & 0.114 & 1.174 \\
		$\fnl=0$, $\sigma_{\rm FP}$ & 0.177 & 0.224 & 1.224  \\ 
		\hline
		$\fnl=100$, $\sigma_{\rm F}$  & 0.110 & 0.110 & 1.081 \\
		$\fnl=100$, $\sigma_{\rm FP}$ & 0.120 & 0.156 & 1.056 \\
		\hline
	\end{tabular}
\end{table}

\section{Halo bias on the Goliat-PNG simulations}
\label{sec:bias}
\subsection{Modeling}
\label{sec:model}

Dark matter halos and galaxies are biased tracers of the underlying matter distribution. At large scales, this can be described by a linear relation between the matter overdensity ($\delta$) and the halo overdensity ($\delta_{\rm halo}$) \citep{PBS,Desjacques}: 

\begin{equation}
    \delta_{\rm halo} = b \cdot \delta\, ,
    \label{eq:bias_relation}
\end{equation}
with $b$ defined as the linear bias. 

Hence, the power spectrum, which is is simply the Fourier 2-point function of $\delta$, is  described as

\begin{equation}
    P_{\rm halo}(k) = b^2\cdot P(k) \, ,
\end{equation}
with $P(k)$ being the dark matter power spectrum, which we will be modelling with the \textsc{camb} linear power spectrum\footnote{\url{https://camb.info/sources/}; \citet{camb}.}. 

Whereas for Gaussian initial conditions, the bias is found to be constant at large scales $b=b_g$, in the presence of local Primordial Non-Gaussianities, the bias has been found to follow \citep{Dalal08, Slosar08} : 

\begin{equation}
    b(k) = b_g + 2\delta_c (b_g - p)\cdot \fnl \cdot \frac{1}{\alpha(k,z)}\, ,
    \label{eq:bk}
\end{equation}
with $b_g$ and $p$ being a priori free parameters. $\alpha(k,z)$ is defined in \autoref{eq:alpha} and has a $k^{2}$ dependence, which will dominate the clustering at large scales. It was originally proposed that $p=1$ \citep[][]{Dalal08} in what is called the {\it Universality Relation} (assuming that halo bias only depends on mass), motivated by the Peak-Background Split theory \citep{PBS,Slosar08}. However, we expect this relation to be insufficient and current evidence indicates that $p$ may depend on mass, redshift or type of tracer in the same way that the linear bias parameter $b_g$ does \citep[][]{Slosar08,Biagetti_2017,Barreira_2020,Barreira_2022}. 
Additionally, in some those works \autoref{eq:bias_relation} is rewritten with $b_\Phi = 2 \delta_c (b - p)$, understood as the response of the halo/galaxy density field to perturbations in the gravitational field $\Phi$. 

\subsection{Fits to the simulations}

\subsubsection{Methods}

We are now prepared to fit the modelling described in \autoref{sec:model} to the halo power spectrum measured in \autoref{sec:sims}, using the variance described in \autoref{sec:variance}.
Since the covariance we are using is diagonal, the $\chi^2$ reads: 

\begin{equation}
\label{eq:chi2}
    \chi^2(\vec \theta) = \sum^{k_{\rm max}}_k \frac{ \Big( P_{\rm halo}(k)-b(k,\vec \theta)^2 P_{\rm lin}(k)\Big)^2}{\sigma(k)^2}\, ,
\end{equation}
where $k_{\rm max}$ is the maximum wave-number considered. We find that our results are very stable against $k_{\rm max}$ variations up to $k_{\rm max}=0.09\Mpch$. Hence, we fix that maximum $k$ for the rest of the analysis. $\vec \theta$ represents the free parameters considered, namely combinations of $b_g$, $p$ and $f_{\rm NL}$.

We then consider the standard Gaussian likelihood
\begin{equation}
    \label{eq:likelihood}
   \mathcal{L}(\vec \theta) \propto {\rm exp}(-\chi^2(\vec \theta)/2)\,
\end{equation}
evaluated on a grid. 
When showing 2D likelihood contours, we consider the 1- or 2-$\sigma$ credible region (c.r.) given by $\Delta \chi^2=1$ or $\Delta \chi^2=4$ with respect to the minimum $\chi^2$. When reporting errors on marginalised parameters, they will be given by the inter-quantile region with the 68\% c.r. (center $\pm$ semi-width). 

\subsubsection{$\fnl=0$ simulations}

\begin{figure}
    \centering
    \includegraphics[width=\linewidth]{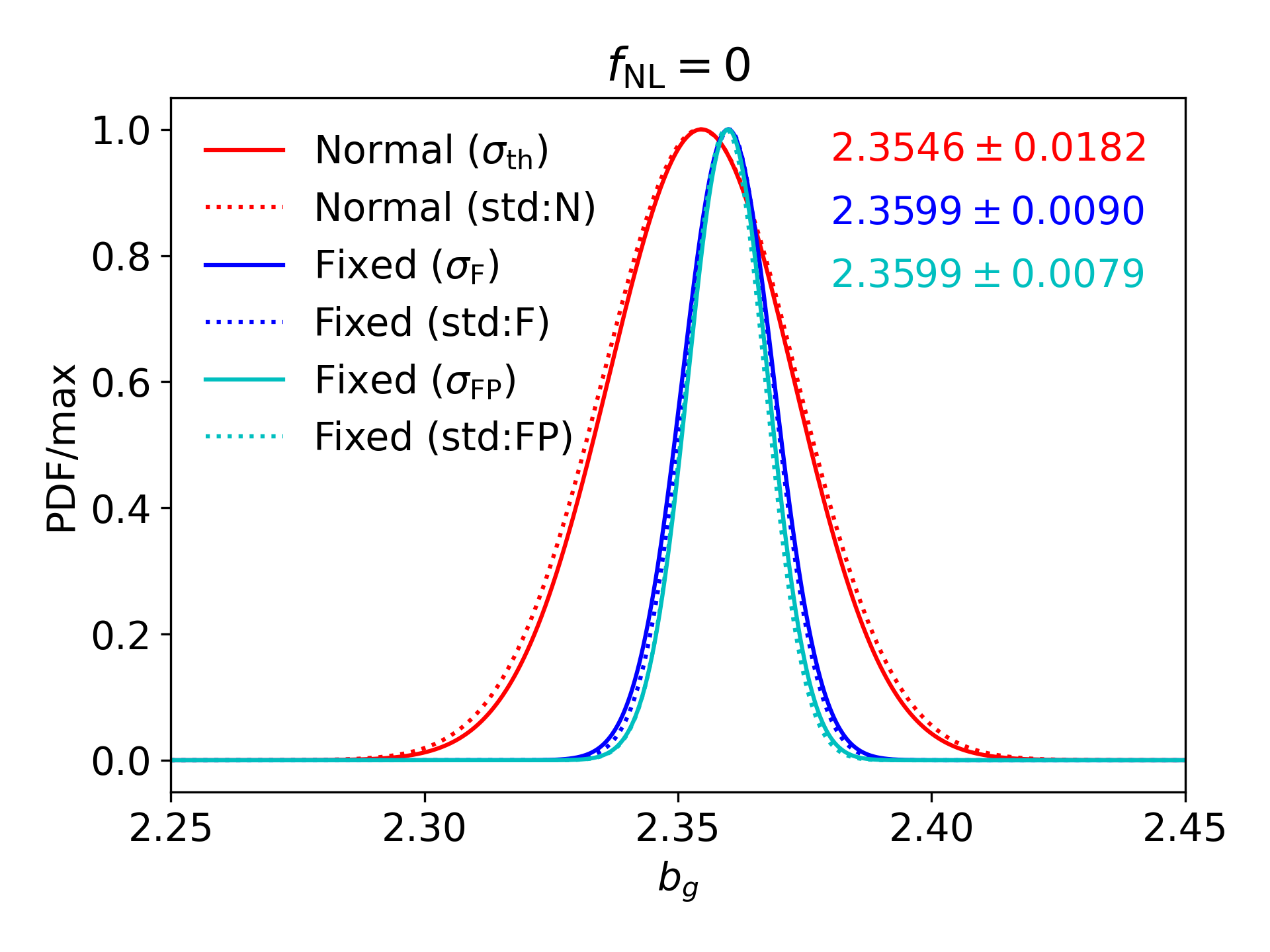}
    \caption{Probability Distribution Function (PDF) for the likelihood (\autoref{eq:chi2}, \autoref{eq:likelihood}) of fitting the measured halo bias on the $\fnl=0$ \gpng\ simulations to a constant $b_g$. By default (solid lines, and reported values) we use the errors, $\sigma$, given by the theoretical/fitting functions described by \autoref{eq:sigma_th} and \autoref{eq:sigma_fix}, whereas the dotted lines represent the PDF derived by using the standard deviation measured in the simulations. The results for the {\it Normal} ({\it Normal}-{\it Original} + {\it Normal}-{\it Inverted}) simulations is shown in red. For the {\it Fixed} ({\it Fixed}-{\it Original} + {\it Fixed}-{\it Inverted}) simulations we consider both the cases of the errors given by only {\it Fixing} (blue) and by both {\it Fixing \& Pairing} (cyan).}
    \label{fig:pdf_fnl0}
\end{figure}
\begin{figure}
    \centering
    \includegraphics[width=\linewidth]{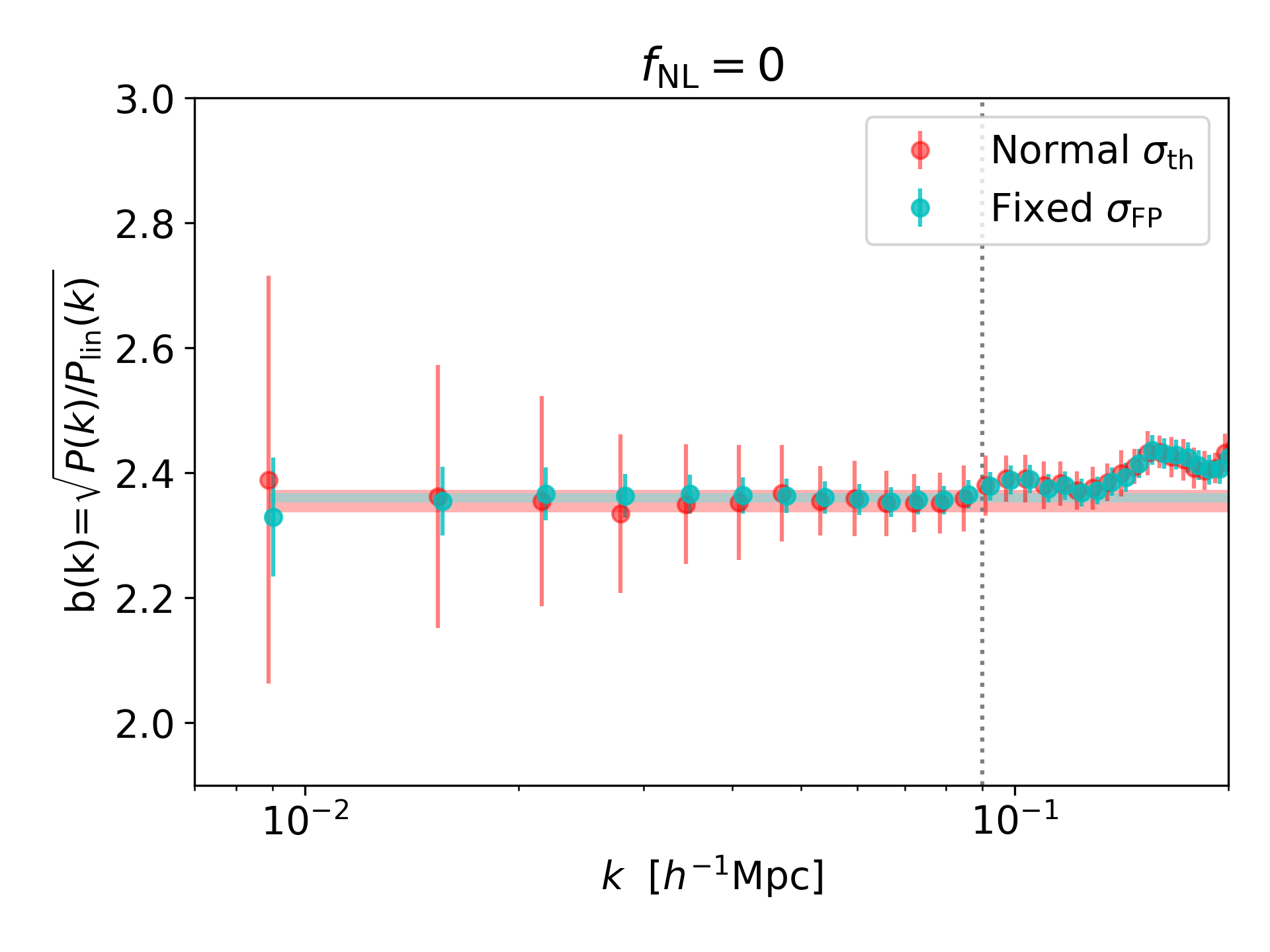}
    \caption{Bias for the $\fnl=0$ simulations. The points represent the halo bias measured from the mean of the simulations ($b(k)=\sqrt{P(k)/P_{\rm lin}(k)}$), with errorbars propagated from 
    the fit in \autoref{eq:sigma_fix}).
    The bands represent the 1-$\sigma$ best fit region for a constant bias $b(k)=b_g$ (values reported in \autoref{fig:pdf_fnl0}), fitted up to $k_{\rm max}=0.09\Mpch$, represented by the vertical dashed line. The cyan points have been slightly shifted side-wise for the purpose of visibility.}
    \label{fig:bias_fnl0}
\end{figure}

We start by fitting a constant bias $b(k)\equiv \sqrt{P(k)/P_{\rm lin}(k)}=b_g$ to the $\fnl=0$ simulations in \autoref{fig:pdf_fnl0}. First, we consider the mean of all the {\it Normal} simulations\footnote{Recall that at this stage the average of all {\it Normal} simulations include already the {\it Original-Inverted}. And similarly for the {\it Fixed}.}, with the expected error of a single simulation, $\sigma_{\rm th}$, given by \autoref{eq:sigma_th} and show the results in solid red. We repeat this process for all the {\it Fixed} simulations considering either the error fitted to the {\it Fixed} case, $\sigma_F$, or the error fitted to the {\it Fixed-Paired} case, $\sigma_{\rm FP}$ (\autoref{eq:sigma_fix}, \autoref{tab:fits}). We find that the error on the bias is reduced by a factor $\sim2$ when {\it Fixing} the initial conditions, whereas adding the {\it Pairing} further reduces the error another $\sim10\%$. In order to make sure that our approximations to the errors $\sigma(k)$ do not introduce any significant changes, we also plot in dotted lines the results when using directly the standard deviation measured in the simulations, finding the differences negligible.

In \autoref{fig:bias_fnl0} we show the bias measured from the simulations when using linear theory as a reference. We find that our constant bias fits (bands) represent a good description up to $k_{\rm max}=0.09\Mpch$. 
In general, we have found a good consistency in our description of the bias for the $\fnl=0$ simulations for the different types of simulations considered ({\it Normal} and {\it Fixed}) and the different error estimations.

\subsubsection{$\fnl=100$ simulations}

\begin{figure*}
    \centering
    \includegraphics[width=0.5\linewidth]{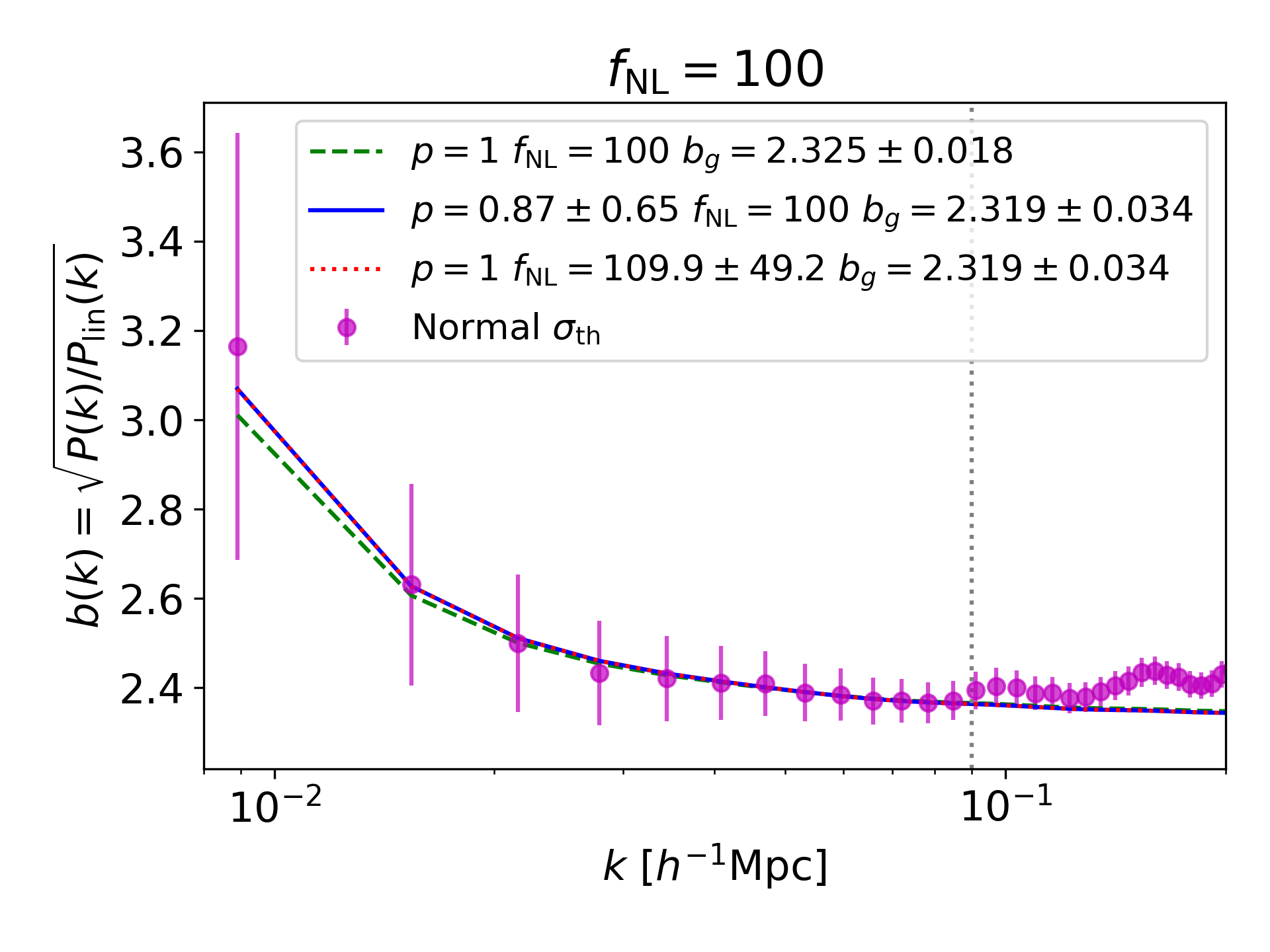}\includegraphics[width=0.5\linewidth]{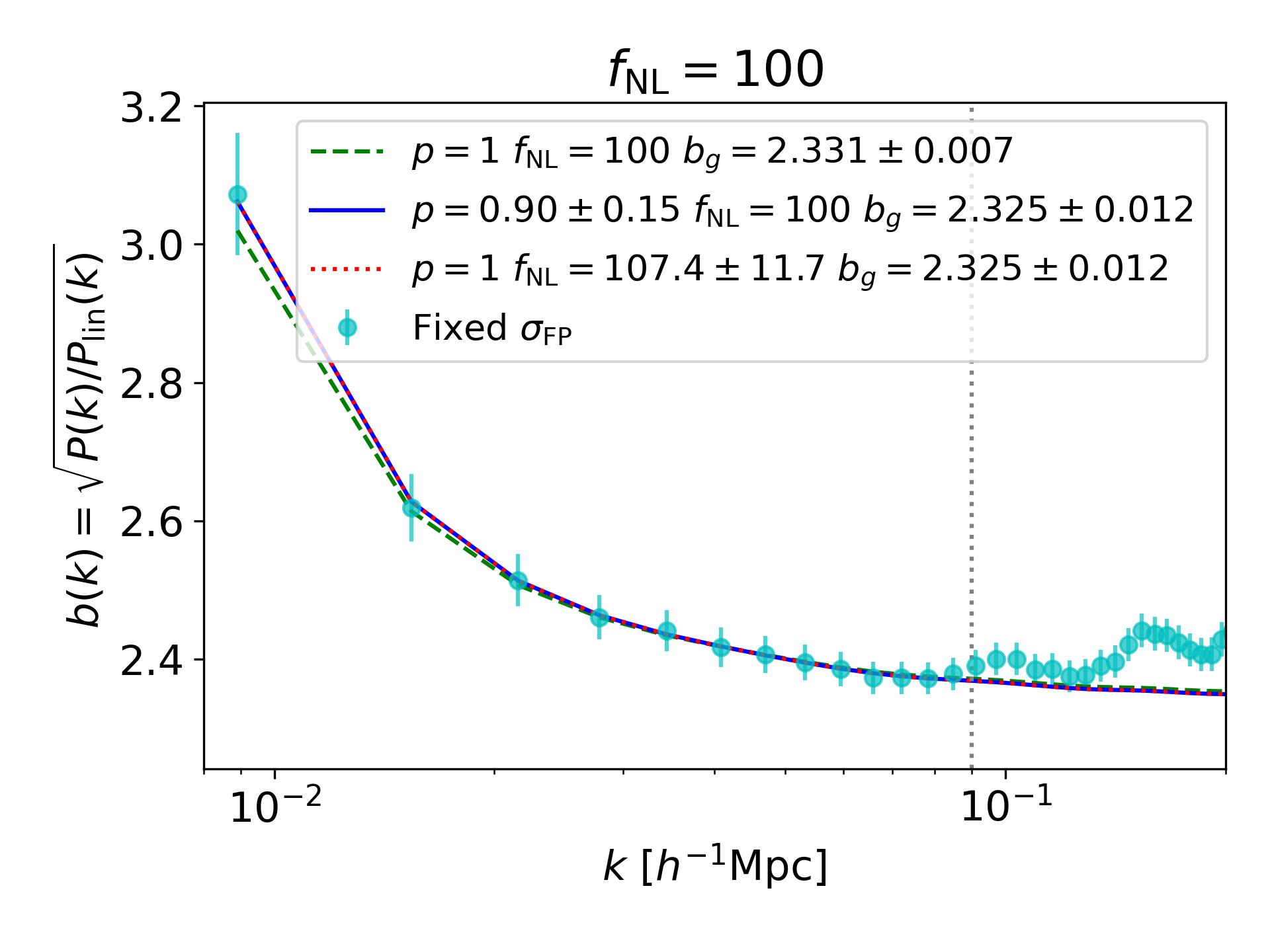}
    \caption{Scale-dependent halo bias derived from the {\it Normal} (left) and the {\it Fixed} (right) simulations. The errorbars represent the measured values from the simulations (error propagated from $\sigma_{\rm th}$, \autoref{eq:sigma_th}, and $\sigma_{\rm FP}$, \autoref{eq:sigma_fix}) and the lines represent the best fits with different subset of the parameters \{$p$, $\fnl$, $b_g$\} set free, following  cases {\it i)}, {\it ii)} and {\it iii)} described in the text. The vertical dotted line represents the maximum $k$ considered ($k_{\rm max}=0.09 h^{-1}{\rm Mpc}$).}
    \label{fig:bias_fnl100}
\end{figure*}

\begin{figure}
    \centering
    \includegraphics[trim=25 0 40 40, clip, width=1.1\linewidth]{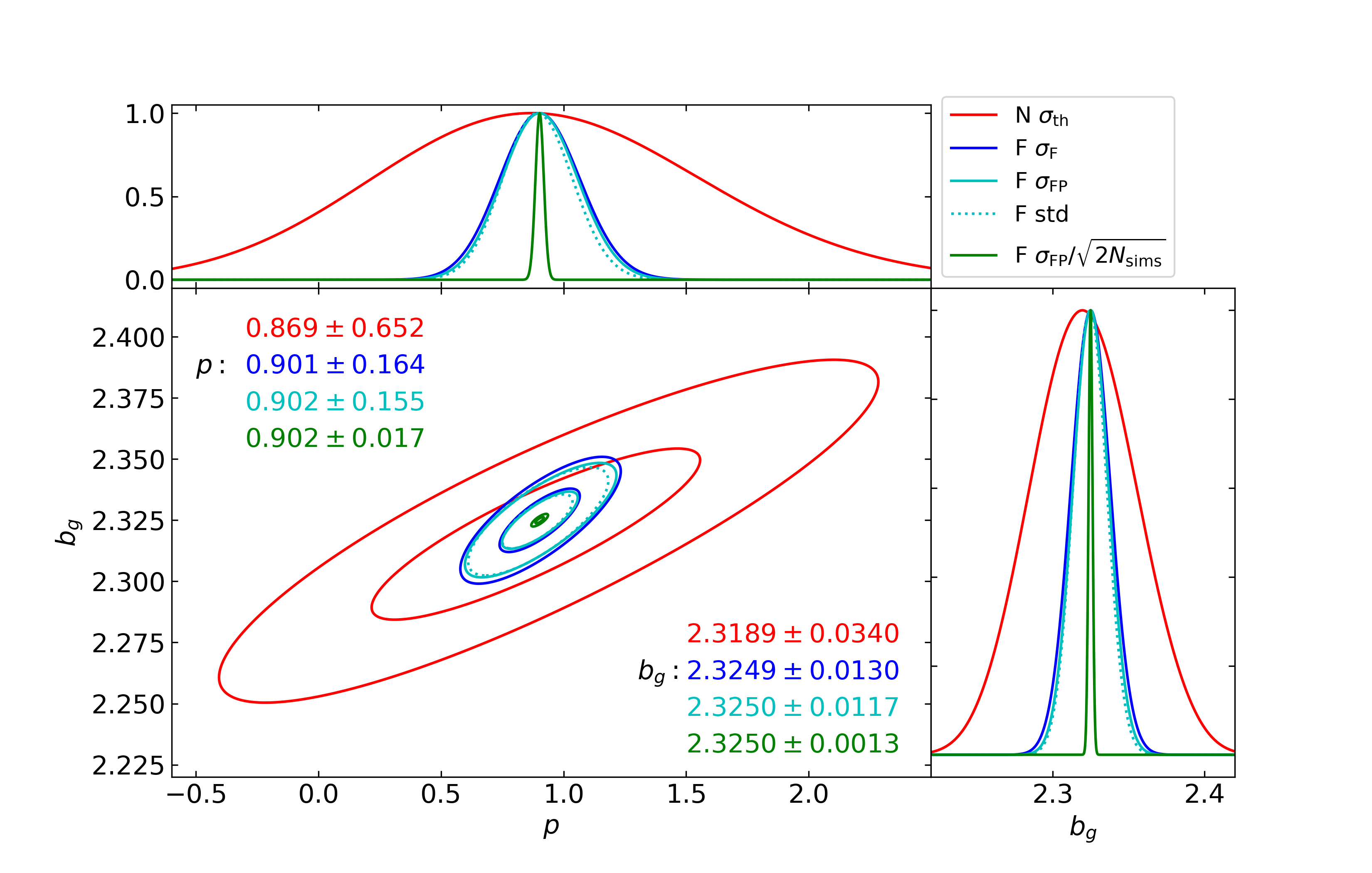}
    \caption{Main panel: 1 and 2-$\sigma$ contours ($\Delta \chi^2=\ 1,\, 4$) of the likelihood in the $\{p,b_g\}$ plane. Top and right panels: marginalised posteriors for $p$ and $b_g$. We show the results for the {\it Normal} simulations with the standard error (red, \autoref{eq:sigma_th}) and the {\it Fixed} simulations with both the {\it Fixed} (blue) and {\it Fixed}-{\it Paired} (cyan) error (\autoref{eq:sigma_fix}, \autoref{tab:fits}). Recall that the {\it Fixed}-{\it Paired} error is compensated with a $\sqrt{2}$ for the volume increment. We also show results for {\it Fixed}-{\it Paired} error directly measure from the standard deviation of the simulations (cyan-dotted). In green, we show the constraints when considering the error on the ensemble of the $2\times41$ simulations. Finally, on the coloured labels we show the 68\% c.r. constraints on $b_g$ and $p$ (we omit the results for the dotted case).
    }
    \label{fig:fit_b_p}
\end{figure}

We perform a similar analysis for the $\fnl=100$ simulations. Here, we consider three fitting procedures:

\begin{itemize}
\renewcommand\labelitemi{}
    \item {\bf {\it i)}} Fixing $p=1$ \& $\fnl=100$, while letting $b_g$ free.
    \item {\bf {\it ii)}} Fixing $\fnl=100$, while letting $b_g$ \& $p$ free.
    \item {\bf {\it iii)}} Fixing $p=1$, while letting $b_g$ \& $\fnl$ free.
\end{itemize}

We show the $b(k)$ measurements from the simulations in \autoref{fig:bias_fnl100}, together with the best fit for the three cases described above. We find that cases {\it ii)} and {\it iii)} give mathematically equivalent results for the $b(k)$ since there is a complete degeneracy between $p$ and $\fnl$ in \autoref{eq:bk}. Precisely, we show both {\it ii)} and {\it iii)} to emphasize that a bias on the assumed $p$ can bias the obtained $\fnl$ and that there is an direct propagation from the error on one to the other one. We also find consistent results between the {\it Normal} and for the {\it Fixed} simulations. 

Case {\it i)} shows slightly worse fit than {\it ii)} and {\it iii)}, as expected since it has one  parameter less. Given that case {\it i)} still shows a good fit and that $p$ is found compatible with 1, one may consider whether it is necessary to consider $p$ different to unity. So far, we have been only considering the error equivalent to a single $V=(1\Gpch)^3$ simulation, as the focus of the paper is to understand what we can infer from single simulations (or pairs), at fixed computational cost. However, for a moment we can consider the error on the ensemble average of the $2\times41$ simulations. This is shown on \autoref{fig:fit_b_p} in green, where we find $p=0.902\pm0.017$, hence, we know that the ensemble of our mocks are best described with $p\neq1$ (implying a deviation from universality, see \autoref{sec:model}), and then, it is worth considering this parameter as free. 

In \autoref{fig:fit_b_p} we also show the constraints on the $\{p,b_g\}$ plane on the {\it Normal} and {\it Fixed} cases, similarly to what was done in \autoref{fig:bias_fnl0} for only $b_g$. We find that {\it Fixing} the initial conditions can reduce the error by a factor $\sim 4$ on $p$ and by a factor $\sim 3$ on $b_g$, when fitted simultaneously for the $\fnl=100$ \gpng\ simulations. Adding the {\it Pairing} gives us an additional $\sim5\%$ and $\sim10\%$ gain, respectively. We checked again that using the standard deviation instead of the fitted errors gives similar results. 

We have found, as motivated in \autoref{sec:intro}, that the gain of using {\it Fixed} initial conditions is huge when we want to constrain the bias parameters ($b_g$, $p$) associated to the PNG halo clustering. {\it Pairing} can give us some mild additional constraint. In the next section, we will study how this gain can also be used in the context of model validation.

\section{Model testing with PNG Fixed, Paired \& Matched simulations}
\label{sec:results}

\begin{figure}
    \centering
    \includegraphics[trim=0 0 0 0, clip, width=0.9\linewidth]{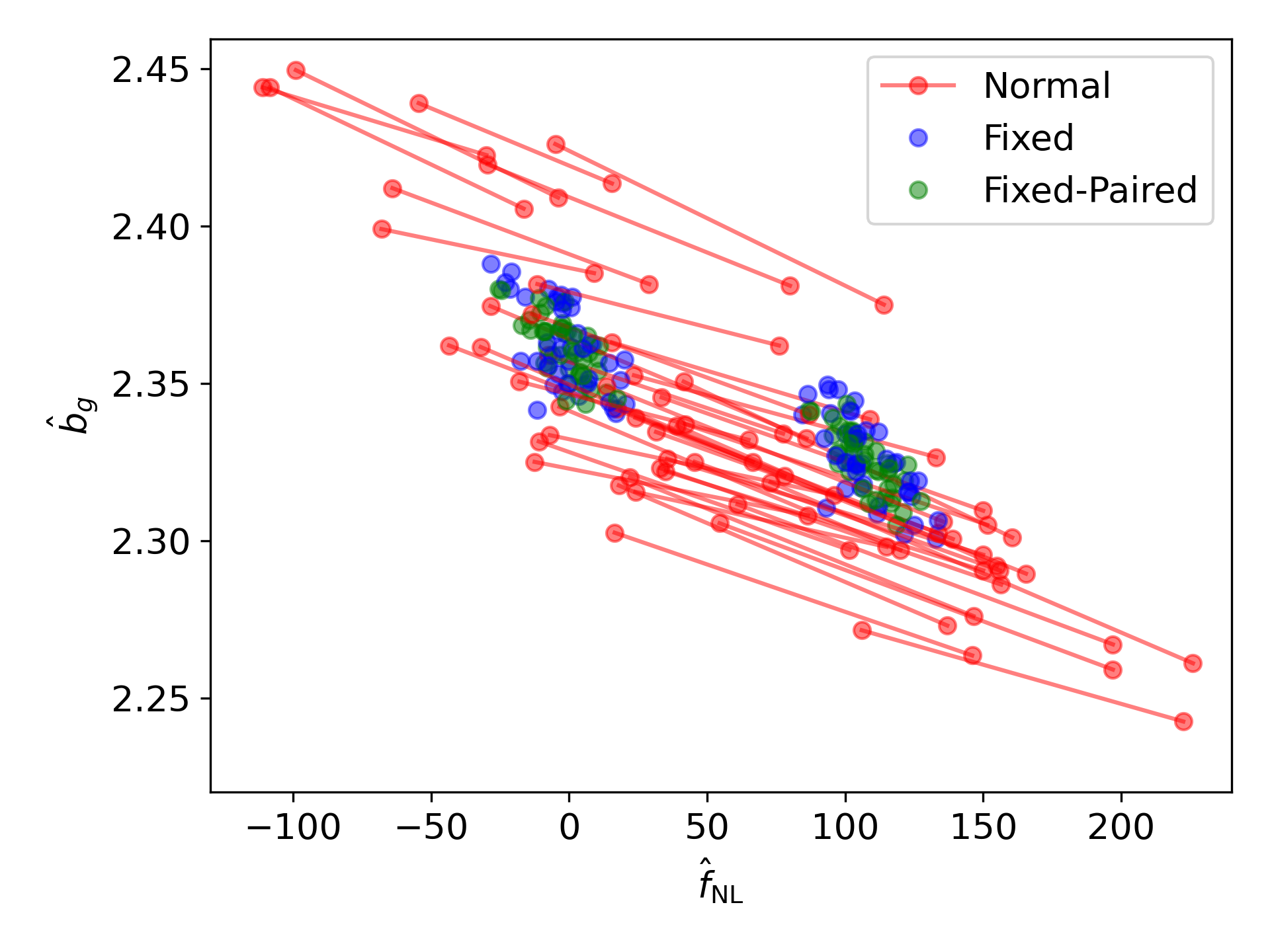}
    \includegraphics[trim=0 0 0 0, clip, width=0.9\linewidth]{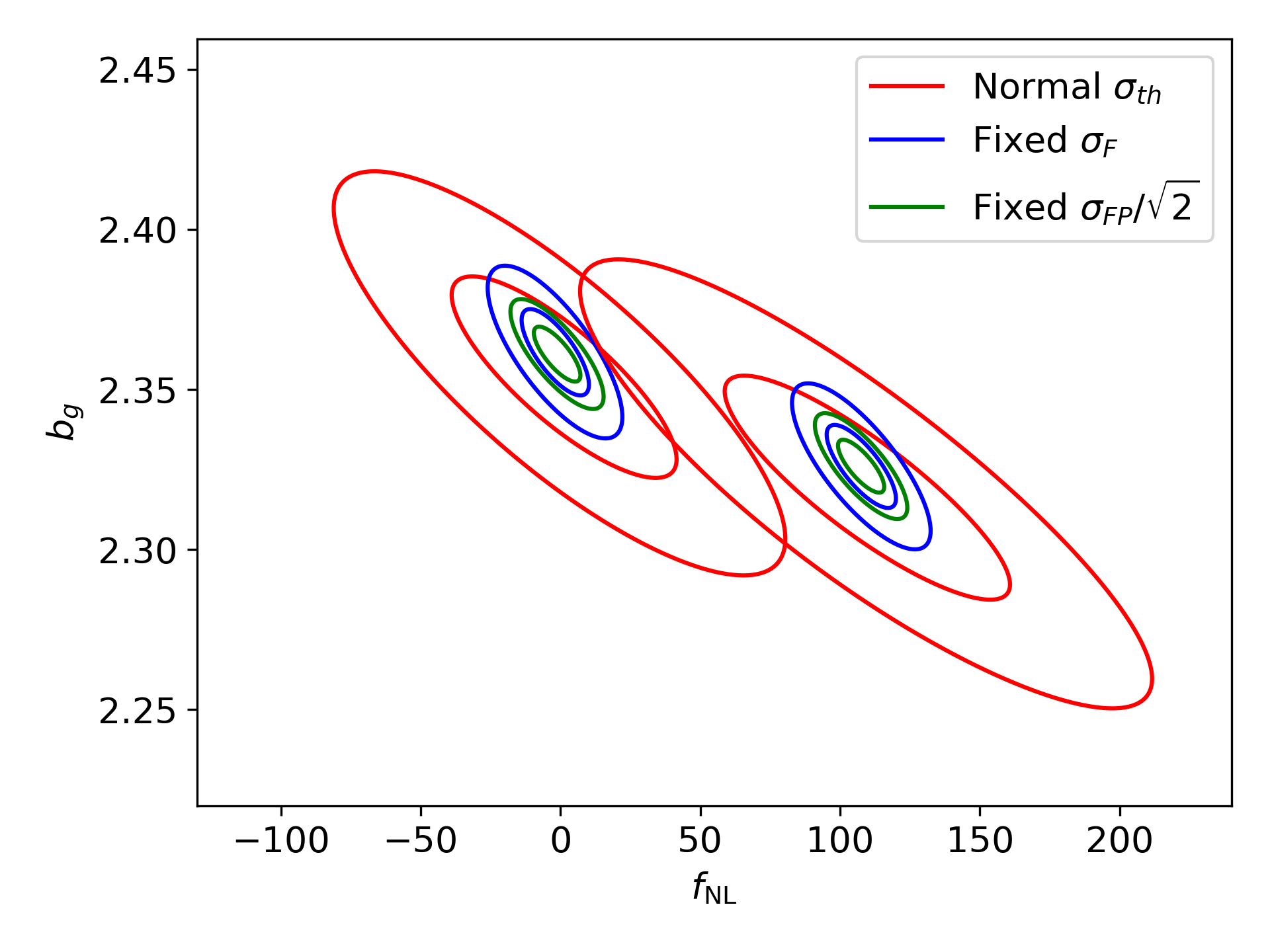}
    \caption{{\bf Top:} We represent the position of the best fits in the $\{\fnl,b_g\}$ plane for all the individual {\it Normal} and {\it Fixed} simulations and also for the {\it Pairs} of {\it Fixed} simulations (using \autoref{eq:pair}). We show the results for both the $\fnl=0$ and $\fnl=100$ simulations. For the {\it Normal} cases we matchwith a red line the simulations using the same stochastic component of the ICs ({\it Matched}), making already very apparent the correlation between the $\fnl=0$ and $\fnl=100$ best fits. {\bf Bottom:} 1-$\sigma$ and 2-$\sigma$ contours of the likelihood on the mean of the simulations. Unlike in \autoref{fig:fit_b_p}, the error used here for the {\it Fixed}-{\it Paired} corresponds to the volume of the two simulations in the pairs. The scatter on the points on the top panel approximately follow the contours shown in the bottom panel. }
    \label{fig:pairs}
\end{figure}

\subsection{Individual fits}

Up to this point, we have always worked with the average of all simulations and their variance. We now turn to analyse the simulations individually or in pairs, in order to show the statistical power of using the {\it Fix}, {\it Pair} and {\it Match} techniques. 
In particular, we will work with $b_g$ and $\fnl$ as free parameters with $p=1$ (case {\it iii)} above) and we choose the peak of the $\fnl$ posterior as our estimator: $\hfnl$. 

Since we are using the same stochastic part of the ICs (the phases and, for {\it Normal} simulations, also the noise in the amplitude, see \autoref{sec:ICs_normal}) for the $\fnl=0$ and the $\fnl=100$ cases, their statistics will be highly correlated. This means that if a single realisation has some large scale $P(k)$ fluctuations that favour a higher measured $\fnl$ for the $\fnl=100$ simulation (from now $\hfnl^{100}$), we should also expect a higher estimation of $\fnl$ for the $\fnl=0$ simulation ($\hfnl^0$), if we run the same analysis for both. This is precisely what we can see in the top panel of \autoref{fig:pairs}. Here, we have fitted individually all the 41 $\fnl=0$ and 41 $\fnl=100$ simulations for the {\it Normal}, the {\it Fixed} and the {\it Fixed}-{\it Paired} cases. For the {\it Normal} cases, we join the simulations with the same initial phases by a line. We find that, whereas the individual fits have a huge scatter (with  the distribution of the $\fnl=0$ and $\fnl=100$ best fits even overlapping), the length and orientation of the magenta line remains relatively constant. 

For the {\it Fixed} and {\it Fixed}-{\it Paired} cases, we find that the scatter in the best fits $\hfnl$ is greatly suppressed, as anticipated. At the bottom panel of \autoref{fig:pairs} we show the 2D constraints derived in the mean of the {\it Normal} and {\it {\it Fixed}} simulations, similar to what was shown in \autoref{fig:fit_b_p} but now on the $\{\fnl,b_g\}$ plane. However, note that here we divide by $\sqrt{2}$ the $\sigma_{\rm FP}$ error as we want to compare it with the scatter of the {\it Fixed} \& {\it Paired} simulations, whose scatter will be intrinsically reduced by the doubling of volume. Qualitatively, the scatter of the best fits in the top panel follows the ellipses shown in the bottom panel. 

\begin{figure*}
    \centering
    \includegraphics[trim=0 0 0 0, clip, width=0.33\linewidth]{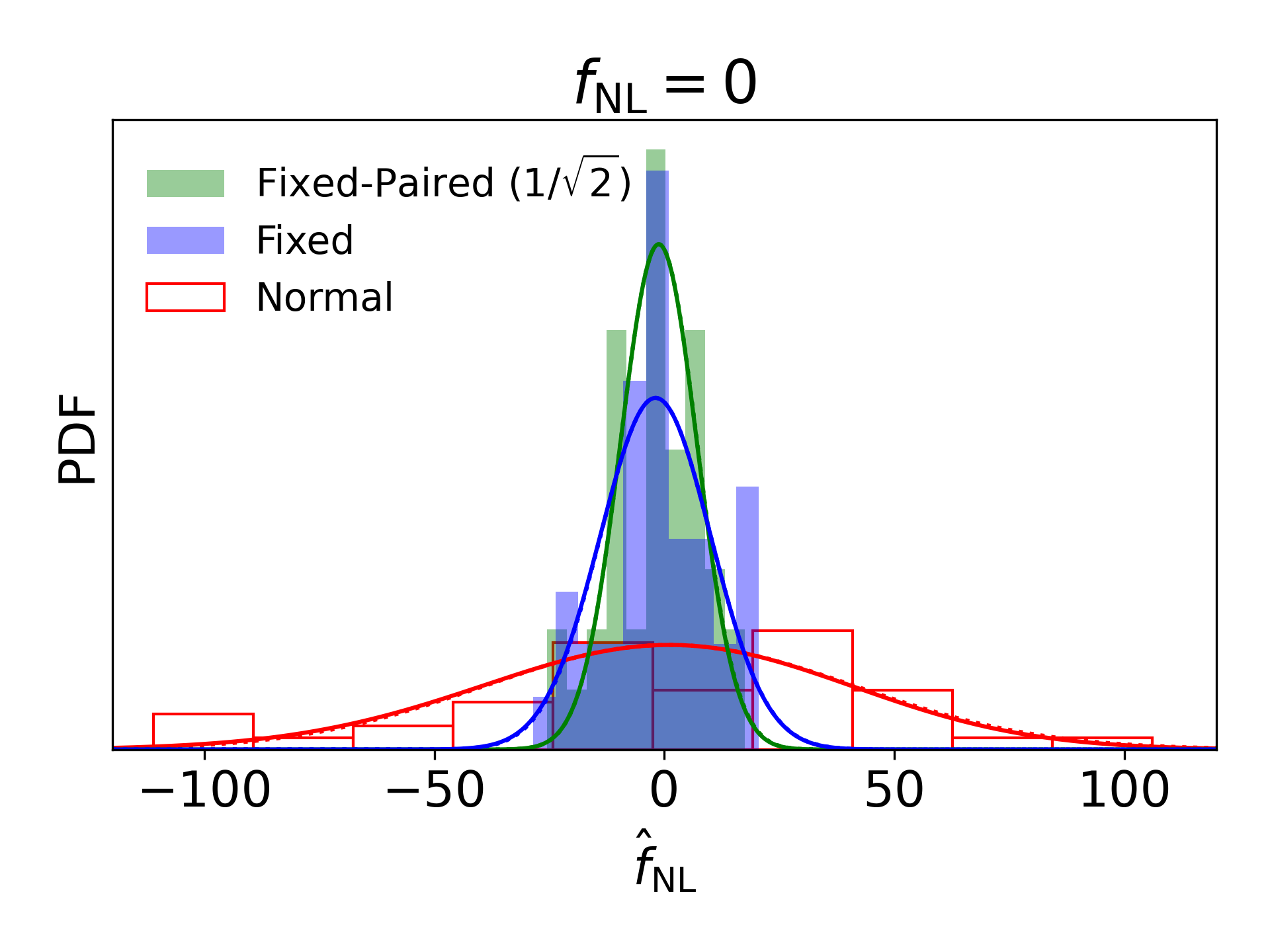}\includegraphics[trim=0 0 0 0, clip, width=0.33\linewidth]{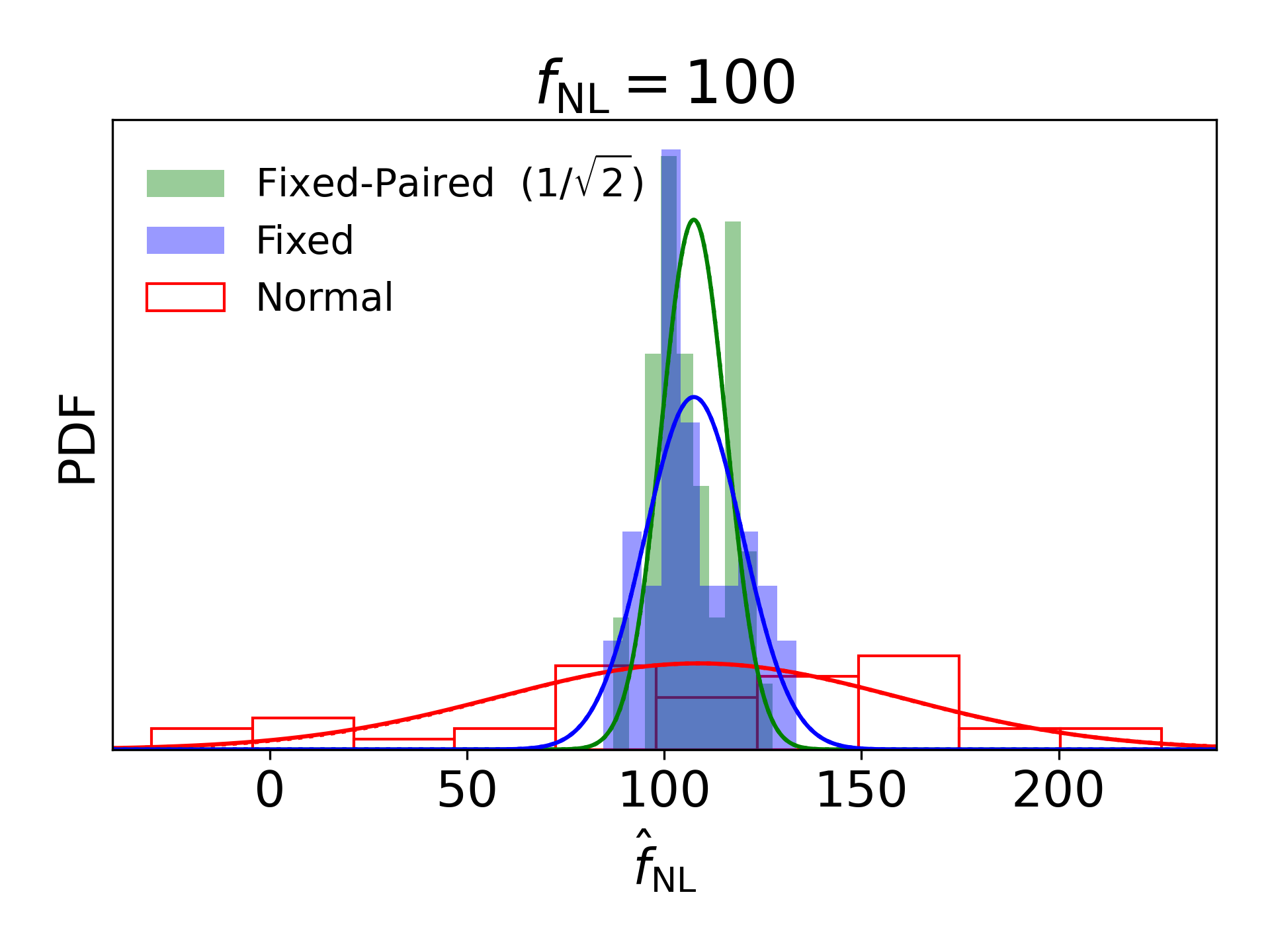}\includegraphics[trim=0 0 0 0, clip, width=0.33\linewidth]{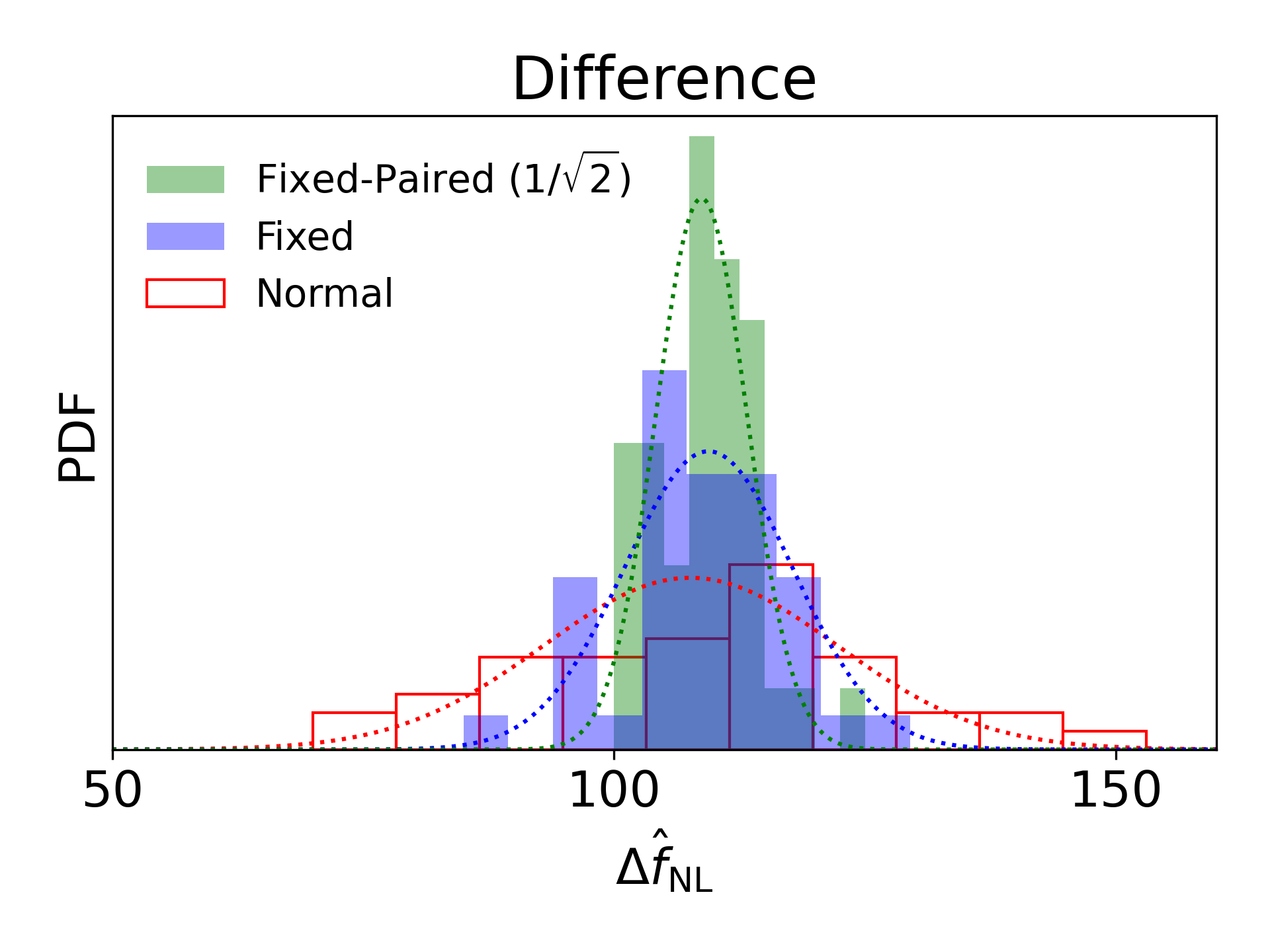}
    \caption{The histograms show the distribution of best fit values of $\fnl$ for the $\fnl=0$ simulations (left), the $\fnl=100$ simulations (centre) and their differences (right). We show on top the posterior on the mean of all the 41 simulations on solid lines, and its Gaussian approximation on dashed lines (barely distinguishable). We find that the difference on the estimated PNG parameter, $\Delta \hfnl$, shows a much more reduced scatter than the individual cases (note the difference of scale on the $x$-axis). We also find that the {\it Fixed} simulations (blue) also greatly reduces the error on $\fnl$ with respect to the {\it Normal} simulations (red). When we consider the {\it Fixed}-{\it Paired} case (green), it further reduces the scatter, but it is compatible with the simple effect of doubling the volume. More details in the text and in \autoref{tab:rho}.}
    \label{fig:PDF}
\end{figure*}

More quantitative results are given on the marginalised results on $\fnl$. The histogram of the distributions of $\hfnl$ for the $\fnl=0$ and $\fnl=100$ simulations is given in the first two panels of \autoref{fig:PDF} and their mean and standard deviation reported in the \autoref{tab:rho} ($1^{\rm st}$ and $3^{\rm rd}$ columns). Next to them, we also show the 68\% c.r. of the posterior on $\fnl$ on the mean of all the simulations (represented by the center and semi-width under the label `Mean'), with the posterior of the mean represented also in \autoref{fig:PDF} as solid lines. We find good consistency between the distribution of the individual fits and the fit on the mean of the mocks (note that we expect a 11\% level of statistical uncertainty on the ${\rm std}$ and 15\% on the mean, given that we are using $N_{\rm sims}=41$ mocks; see \autoref{eq:errerr}). 

When using {\it Fixed} initial conditions, we find a drastic ($\times \sim 4$) reduction in the uncertainty (and scatter) of $\fnl$ with respect to the {\it Normal} simulations. However, when also {\it Pairing} them, we do not find evidence for an improvement on $\sigma(\fnl)$ beyond the one expected by the increment of volume: whereas the error on the mean does show some slight improvement, the measured scatter increases (when corrected by the $\sqrt{2}$ factor). 
Nonetheless, this subtle differences fall within the expected statistical differences by having a limited number of mocks. 
Additionally, for non-Gaussian likelihoods, the scatter on the peak of the posterior may not coincide with the $\sigma$ derived on the mean $P(k)$. These subtleties are beyond the scope of this paper as we focus on the large improvements obtained by the {\it Fixing} and by the usage of correlation between {\it Matched}-ICs boxes (see below). Hence, we simply conclude that {\it Pairing} does not add a significant improvement to the measurement of $\fnl$ with our analysis pipeline.

\subsection{Correlating Matched ICs}

\begin{table*}
	\centering
	\caption{Summary statistics of individual and global fits to the \gpng\ simulations: {\bf 1)} mean and standard deviation of the individual  best fits $\hfnl$ for the $\fnl=0$ simulations; {\bf 2)} 1-$\sigma$ c.r. from the fit on the mean of the $\fnl=0$ simulations and used for its Gaussian approximation; {\bf 3)} mean and standard deviation of the individual $\hfnl$ best fits for the $\fnl=100$ simulations; {\bf 4)} 1-$\sigma$ c.r. inferred from the fit on the mean of the $\fnl=100$ simulations; {\bf 5)} gain in effective volume with respect to a standard simulation due to the type of initial conditions based on the ${\rm std}$ [and based on the $\sigma$ on the mean] on the $\fnl=100$ simulations; {\bf 6)} Pearson correlation coefficient between $\hfnlz$ and $\hfnlh$; {\bf 7)} mean and standard deviation of the {\it Matched} difference between $\hfnlz$ and $\hfnlh$; {\bf 8)} derived $\mu$ and $\sigma$ parameters under the Gaussian approximation for $\Delta \fnl$; {\bf 9)} total gain in effective volume with respect to a standard non-{\it Matched} simulation due to both the use of {\it Matched}-ICs and the type of initial conditions based on the std [and based on the Gaussian $\sigma$] (with $V_{\rm eff}\propto 1/\sigma^2$). Each of the three first rows represents a different case of initial conditions ({\it Normal}, {\it Fixed} or {\it Fixed-Paired}), whereas in the last row we multiply the errors of the {\it Fixed-Paired} by $\sqrt{2}$ to compensate by the doubling of volume. }
	\label{tab:rho}
	\begin{tabular}{lrrrrrrrrr} 
		\hline
		  & $\langle \hfnlz \rangle \pm {\rm std}(\hfnlz)$ & Mean  & $\langle \hfnlh \rangle \pm {\rm std}(\hfnlh)$ & Mean & $V_{\rm eff}/V$ & $\rho$ & $\langle \Delta \hfnl \rangle \pm {\rm std}(\Delta \hfnl)$ & Gauss. & $V_{\rm eff}/V$ \\
		\hline
		{\it Normal}    & $1.9\pm47.0$ &$1.1\pm40.3$  & $111.4\pm60.5$ & $108.7\pm50.7$ & 1 [1] & 0.97 & $109.5\pm18.1$ & $107.6\pm14.6$ & 11 [12] \\
		{\it Fixed}         & $-1.3\pm11.8$ & $-1.9\pm12.0$ & $107.6\pm12.1$ & $107.5\pm12.4$  & 25 [17] & 0.76 & $108.9\pm8.3$ & $109.4\pm8.4$ & 53 [36] \\
		{\it F.}-{\it Paired}  & $-1.8\pm9.5$ & $-1.2\pm8.3$ & $107.4\pm9.3$ & $107.5\pm8.2$ & -- & 0.85 & $109.2\pm5.2$ & $108.7\pm4.6$ & -- \\
		$\, \, \times \sqrt{2}$  & $\pm13.5$ &    $\pm11.8$ &    $\pm13.2$ &    $\pm11.7$ & 21 [19] & -- &    $\pm7.3$ &    $\pm6.4$ & 89 [3] \\
		\hline
	\end{tabular}
\end{table*}

Motivated by the top panel of \autoref{fig:pairs}, we can now consider our new estimator as the difference between the best fit value of $\fnl=100$ and the $\fnl=0$ simulations:

\begin{equation}
    \Delta \hfnl = \hfnl^{100} - \hfnl^0\ .
    \label{eq:diff}
\end{equation}

Then, the propagated error on this estimator, taking into account the Pearson correlation coefficient $\rho$ between $\hfnlz$ and $\hfnlh$, is

\begin{equation}
    \sigma(\Delta \hfnl)^2 =  \sigma(\hfnlh)^2 + \sigma(\hfnlz)^2 - 2\cdot \rho \cdot \sigma(\hfnlh) \sigma(\hfnlz)\, . 
    \label{eq:correlation}
\end{equation}

For this purpose, we measure the correlation coefficients between the best fits of the $\fnl=0$ and $\fnl=100$ simulations for the {\it Normal}, {\it Fixed} and {\it Fixed}-{\it Paired} cases, shown in \autoref{tab:rho}.

On the right panel of \autoref{fig:PDF} we show the histogram of the  difference of best fits, $\Delta\hfnl$. Indeed, we find that the scatter of $\Delta\hfnl$ is greatly reduce if we compare it with the $\fnl=100$ (or $\fnl=0$) case (note the change of scale in the $x$-axis). 
We also infer a Gaussian PDF with the variance given by \autoref{eq:correlation} (inputting the $\sigma$s derived for the mean of all the $\fnl=0$ $\fnl=100$ simulations) and the mean given by the difference of the peak of the posterior represented in the left and center panels. 
We find a good agreement between the errors expected by \autoref{eq:correlation} and the distribution of $\Delta \hfnl$, as reported in \autoref{tab:rho}.

We find a significant gain in precision, when using the correlation of phases between two simulations with {\it Matched} initial conditions. This can be used in general to validate a model or to constrain nuisance parameters associated with it. 
The larger the correlation, the larger the gain we can obtain from using this technique. Hence, this gain is larger for the {\it Normal} simulations, then the {\it Fixed}-{\it Paired}, and finally the {\it Fixed}. This larger correlation for the {\it Normal} simulations is expected, since much of this correlation comes from having the same cosmic variance realisation at large scales. When {\it Fixing} the ICs, we reduce much the cosmic variance influence, thus, reducing this correlation. Nevertheless, the {\it Matched} {\it Fixed} simulations still present smaller scatter (${\rm std}(\fnl)\approx 8$) than the {\it Normal} simulation (${\rm std}(\fnl)\approx 18$). Hence, these two techniques can be used together to gain even more information. 
For the {\it Fixed}-{\it Paired} case, given the larger correlation than in the {\it Fixed} case, we see a further reduction in the uncertainties. In this occasion, both estimators of the {\it Fixed}-{\it Paired}  error (std and the Gaussian $\sigma$), go in the direction of having a more precise measurement of $\fnl$ (after compensating for the $\sqrt{2}$ factor). However, this gain is still not statistically significant for the {\it Fixed}-{\it Paired} case, given the uncertainty on the standard deviation ($11\%$). 

\subsection{Model validation}

With the {\it Fixed}-{\it Paired} simulations and using the {\it Mathced}-ICs correlation, we obtain a value of $\langle \Delta \hfnl \rangle= 109.2\pm5.2$ (or $\Delta \hfnl = 108.7\pm4.6$ for the Gaussian approximation), which is marginally in tension (almost $2\sigma$) with the known $\Delta \fnl=100$ of our simulations. 
This can be interpreted as a hint that our model ($p=1$) is inaccurate, as we know it is the case: we found in \autoref{fig:fit_b_p} that when combining all the $2\times41$ boxes we obtain $p=0.902\pm0.017$.

Otherwise said, a single pair of $\fnl=100$ {\it Fixed} \& {\it Paired} $N$-Body simulations with the properties of \gpng\ (mainly $L=1\Gpch$) {\it Matched} to another {\it Fixed-Pair} with $\fnl=0$ with the same initial phases is expected to detect a $\sim$2-$\sigma$ discrepancy with the halo clustering predicted by the universality relation ($p=1$). We remind the reader that this relation is expected to break (see \autoref{sec:model}). In order to achieve the same level of accuracy ($\sigma(\fnl)=5$) with {\it Normal} non-{\it Matched} simulations (single error of $\sigma(\fnl)=60$), we would need $\sim 144$ individual simulations (as the error scales as $\sigma\propto1/\sqrt{V}$). 

If we redo our analysis with $p=0.902$, we obtain $\langle \Delta \hfnl \rangle=101.7 \pm 4.8$ (from the {\it Fixed}, {\it Paired} \& {\it Matched}  mocks), an unbiased result. 
Another approach is to re-interpret the measured $\Delta \hfnl$ in terms of $p$. Arguably, what we are able to measure is the joint $(b_g-p)\fnl$ factor ($\propto b_\Phi\cdot \fnl$). Then, the we can estimate $p$ as

\begin{equation}
    \hat p = \hat b_g - (\hat b_g -1)\frac{\hfnl}{\fnl^{\rm true}}\, ,
    \label{eq:p}
\end{equation}
recovering $\hat p = 0.88$ when inputting $\hfnl^{100}=109.2$ retrieved from the {\it Fixed}, {\it Paired} \& {\it Matched}  mocks. Alternatively, if we apply \autoref{eq:p} with $\fnl^{\rm true}=100$ to each mock individually, we obtain $\langle \hat p \rangle \pm {\rm std}(\hat p)= 0.87\pm0.81$, $0.90\pm0.16$, $0.90\pm0.12$  for the {\it Normal}, {\it Fixed} ad {\it Fixed}-{\it Paired} cases, respectively, when inputting $\hfnl=\hfnl^{100}$ (non-{\it Matched}) and $\langle \hat p \rangle \pm {\rm std}(\hat p)= 0.88\pm0.24$, $0.88\pm0.11$, $0.88\pm0.07$, when inputting $\hfnl=\Delta \hfnl$ (i.e. {\it Matching}). For the {\it Matched} + {\it Fixed}-{\it Paired} analysis, we  retrieve again a 2-$\sigma$ hint against the universal relation model $p=1$.

\section{Summary}
\label{sec:summary}

In this work, we have studied combinations of three techniques designed to reduce the variance associated to cosmological simulations by tweaking the Initial Conditions (ICs, \autoref{sec:ICs}): the {\it Fix} \& {\it Pair} techniques \citep[proposed in ][]{Angulo16} and the {\it Matching}. First, {\it Fixing} removes the variance input to the clustering of the of the ICs, greatly reducing the variance at the late times as well. Secondly, {\it Pairing} consists on running a second simulation with the phases {\it Inverted} on the ICs (and the rest of the setup unchanged), which is able to cancel out some contributions to the variance, when combined with the {\it Fixing}. Lastly, the {\it Matching} consists of running simulations with different cosmologies but the same random realisations of the stochastic part of the ICs (for the phase and amplitude). With {\it Matched}-ICs, the retrieved clustering statistics are correlated for the different cosmologies and part of the noise can be canceled out. Whereas the three techniques have been used in the past in a qualitative/implicit way, here we have proposed a framework to utilise (combinations of) these three techniques to increase significantly the precision retrieved from simulations in a quantitative and explicit way. In particular, we put the focus on the usage of simulations to validate galaxy/halo clustering models and also to put constraints/priors of nuisance (bias) parameters associated with them. We note that a recent work by \citet{Zennaro21} already uses {\it Fixed} \& {\it Paired} simulations to constrain (Gaussian) bias parameters down to a reduced uncertainty, with the methodology just realeased in \citet{Maion22}.

In order to illustrate the potential of these techniques, we focused on constraining local Primordial Non-Gaussianities (PNG) with the halo power spectrum $P(k)$. PNG is parametrised by $\fnl$ and it induces a scale-dependent bias at very large scales, where the {\it Fix} technique is specially powerful in reducing the variance. For that purpose, we run the \gpng\ suite: a set of $328$ ($41\times2\times4$) $N$-body simulations, for which we have $41$ different initial random seeds for 2 different values of $\fnl$, $0$ \& $100$, and 4 types of ICs: the {\it Normal}-{\it Original}, the {\it Normal}-{\it Inverted}, {\it Fixed}-{\it Original} and {\it Fixed}-{\it Inverted} (\autoref{sec:sims}).

Our first goal was to validate the usage of the {\it Fix} technique with local-PNG, since {\it Fixing} already induces a type of non-Gaussianity (the Rayleigh PDF is substituted by a Dirac delta). In \autoref{fig:Pk} we find that the $P(k)$ ratio of the {\it Fixed} to {\it Normal} simulations is nearly identical for $\fnl=0$ and $\fnl=100$ and, in both cases, within the noise level. Additionally, in the Appendix \ref{app:validation}, we include a series of  tests that validate the usage of {\it Fixed}-PNG simulations not only for halo power spectrum, but also for the halo bispectrum as well as for the dark matter power spectrum and bispectrum of both initital conditions and late time ($z=1$) snapshots. Additionally, we also verify that the  {\it Fixed}-PNG simulations result unbiased when studying higher resolutions or larger volumes. 

On a second step, we quantified in \autoref{fig:sigma} the reduction on the $P(k)$ variance ($\sigma(k)^2$) introduced by the {\it Fix} and {\it Pair} techniques. Again, the results are found very similar for the $\fnl=0$ and the $\fnl=100$ cases. The {\it Fixed} shows a scale-dependent variance reduction similar to \citet{Unitsim}, \citet{DESIsim2} or \citet{Maion22}, which we fit here with a smoothed step function (\autoref{eq:sigma_fix}) that we use for the rest of the paper. When adding the {\it Pairing} to the {\it Normal} simulations, we actually obtain an increment on the variance at large $k$ in line with \citet{Unitsim}, which goes in opposite  direction to the original motivation. However, when adding the {\it Pairing} to the {\it Fixing} we find slight reduction in the variance at intermediate scales ($k\sim 0.1 \Mpch$), which is in line with the original proposal in \citet{Angulo16}.

In \autoref{sec:bias}, we studied the halo bias of the \gpng\ and validated our modelling and fitting pipeline. We found that we can describe well the halo clustering with a linear theory + linear bias up to $k=0.09 \Mpch$ for our simulations. For that, we use a constant bias for $\fnl=0$ and a the PNG scale-dependent bias (\autoref{eq:bk}) for $\fnl=100$. When considering the ensemble of the $2\times41$ $\fnl=100$ {\it Fixed} \& {\it Paired} simulations we find that the clustering predicted by the universal mass relation \citep[][\autoref{eq:bk} with $p=1$]{Dalal08} is insufficient (as expected) and we obtain $p=0.902\pm0.017$.

Finally, in \autoref{sec:results} we set up the framework to validate galaxy/halo clustering models with combinations of the {\it Fix}, {\it Pair} and {\it Match} techniques. We fitted the bias and PNG parameters $\{b,\fnl\}$ individually for each the simulations (or pairs) for the {\it Normal}, {\it Fixed} and {\it Fixed}-{\it Paired} cases. We find that $P(k)$ variance reduction due to the {\it Fixing} results in a reduction on $\sigma(\fnl)$ by a factor of $\sim 4-5$ (equivalent to gaining a factor of $\sim20$ in simulated volume) for the $\fnl=100$ simulations. Adding the {\it Pairing} does not show a significant gain beyond the doubling of volume. 

The ICs {\it Matching} between the $\fnl=0$ and the $\fnl=100$ simulations results in a high correlation on the fitted $\fnl$,  $\hfnl^0$ and $\hfnl^{100}$ ($\rho=0.97$, $0.76$, $0.85$, for the {\it Normal}, {\it Fixed} and {\it Fixed}-{\it Paired}, respectively). 
Using explicitly their correlation coefficients, we defined a new variable $\Delta \hfnl =\hfnl^{100}-\hfnl^0$ (\autoref{eq:diff}), whose variance is greatly reduced. This allows us to constrain $\Delta \fnl$ with a precision increased by an additional factor of $\sim$ 3--4 for the {\it Normal}, $\sim$ 1.5 for the {\it Fixed} and $\sim$2 for the {\it Fixed}-{\it Paired} (we have multiplied this value by $\sqrt{2}$ to compensate the doubling of volume) cases with respect to their equivalent non-{\it Matched} simulations.

By combining the {\it Fix}, {\it Pair} and {\it Match} techniques altogether we  inferred an uncertainty on one pair of simulations of $\sigma(\fnl)= 5$, whereas one single simulation with $\fnl=100$ yields an uncertainty of $\sigma(\fnl)= 60$. We would have needed $144$ simulations to reach the same level of precision. 

In terms of model validation, combining the {\it Fix}, {\it Pair} and {\it Match} techniques allows us to find a $\sim2\sigma$ hint of deviation from the prediction from \citet{Dalal08} ($p=1$) with one single pair: $\hfnl=109\pm5$ or, equivalently, $p=0.88\pm0.07$ (recall that the ensemble of simulations gave us the reference value of $p=0.902\pm0.017$). Thus, we have shown that explicitly using the variance reduction of {\it Fixing}, {\it Pairing} and {\it Matching} can allow us to test models down to a much smaller error on the cosmological parameters ($\fnl$ in our case) and, therefore, to detect inaccuracies in the clustering model that would have otherwise been undetected. Alternatively, these techniques can also be used to constrain nuisance/bias parameters ($p$ in our case) to a much higher precision. These constraints can later be used to construct informative priors to constrain the data. This is particularly important for PNG analysis with the 2-point statistics, since $\fnl$ and $p$ are completely degenerated.

\section{Outlook}
\label{sec:outlook}

We have shown that explicitly using the variance reduction of {\it Fixed}, {\it Paired} and {\it Matched} simulations can lead us to a better understanding  of halo/galaxy clustering models. In particular, these simulations can be used to validate down to an increased accuracy the analysis tools that we intend to implement in data. Alternatively, we can use these simulations to tighten the priors on nuisance parameters such us the PNG response $p$ studied here or other bias parameters, such as the ones studied in \citet{Zennaro21}. If one only wanted to put priors on bias parameters, an alternative approach is to use the simulation dark matter power spectrum as our theory \citep[as done in][]{Zennaro21}, this would likely capture most of the information that we are here capturing with the {\it Matching}. However, in this paper we put the focus on the validation a galaxy clustering models that we would plan to use on data, hence our theory may not rely on information that we are not able to retrieve from the data (such as the matter power spectrum).

Constraining PNG is one of the main goals for surveys such as DESI\footnote{\url{https://www.desi.lbl.gov/}}, Euclid\footnote{\url{https://sci.esa.int/web/euclid}} or SphereX and one of the main motivations for an intensity mapping program on the SKAO  \citep{DESI,DESI_white,Euclid_white,SphereX_cosmo,SphereX_cosmo_cross,SKA_cosmo}. However, without any prior on $p$ (\autoref{eq:bk}, or $b_\phi$, as done in \citealt{Barreira_2022}), the PNG constraints are completely degenerated with it (as we displayed in \autoref{fig:bias_fnl100}). Hence, building simulation efforts to put tight and robust priors on the PNG bias parameters is a necessary step to measure accurate and precise constraints on $\fnl$ by using the LSS. With the methodology proposed here we have shown that a single {\it Fixed}-{\it Pair} of $V=1[\Gpch]^3$ simulations with $\fnl=100$ {\it Matched} to an existing $\fnl=0$ {\it Fixed}-{\it Pair} can be used to test the PNG halo clustering modelling down to an uncertainty of $\sigma(\fnl)=5$, comparable with current CMB constraints and upcoming constraints from Euclid or DESI \citep{Euclid_white,DESI}. This implies a reduction of the computational costs invested in simulations by a factor $\sim 140$ with respect to using {\it Normal} non-{\it Matched} simulations. 

The techniques proposed here can be used with existing {\it Fixed} and {\it Paired} simulation suits such as UNITsims \citep{Unitsim}, BACCO \citep{Bacco} or Quijote \citep{Quijote} and possible extensions of them. An additional advantage of using {\it Fix}, {\it Pair} and {\it Match} together is that one can use a high mass resolution in order to resolve well the halos of interests for future surveys (log$M_h\sim11$ for Euclid and DESI) in a $V\sim 1 ({\rm Gpc}/h)^{3}$ box with reasonable computing resources ($N=4096^3$ in UNITsim), while keeping a reduced uncertainty on $\fnl$ (or $p$ or $b_\phi$). This way, one could also use semi-analytical models of galaxy formation \citep[as recently done in ][ to populate UNITsim with H-$\alpha$ galaxies for Roman/Euclid]{Zhai21,Knebe22} to understand the effect of galaxy formation on $p$ or $b_\phi$ for different tracers \citep[as done in][]{Barreira_2020,Barreira21}.

One caveat could be the need of many simulations to measure on the one hand the variance reduction induced by the {\it Fixing} and {\it Pairing} (which \citealt{Zhao21} showed that are very sample dependent) and, on the other hand the correlation coefficient associated to the {\it Matching}. However, these can be computed in fast simulations requiring less computational resources such as COLA \citep{cola}, FastPM \citep{fastpm}, HALOGEN \citep{halogen} or EZmocks \citep{EZmocks}.

We envision this paper as a first step into building a suite of simulations specifically aimed at understanding the bias parameters associated to PNG well enough to put informative and robust priors on them. Simultaneously, this set of simulations will also serve to validate PNG analysis pipelines and for preparation of upcoming LSS surveys (particularly, DESI and Euclid), not only by using the halo power spectrum, but also other observables such as the Halo Mass Function, the galaxy bispectrum, the probability distribution functions, zero-bias tracers, etc \citep{MVJ00,LoVerde,Cabass_2022,Friedrich20,Castorina18}. Additionally, we expect that part of the methodology used here can be used to validate galaxy clustering models beyond PNG analysis and to enhance the statistical power of cosmological simulations in general.

\section*{Acknowledgements}

{\it Author contributions}: SA led the project and analysis and wrote the paper. AGA made several important contributions, running great part of the simulations and the validation performed in the appendix.

We thank Mike S. Wang and Raul E. Angulo for detailed feedback on the paper. We also thank Chia-Hsun Chuang and Walter Riquelme for useful discussions and/or feedback on the paper.

SA was funded by the Atracci\'{o}n de Talento Contract no. 2019-T1/TIC-12702 granted by the Comunidad de Madrid in Spain.  
The simulations were run on the HPC-clusters at the IFT-UAM/CSIC: HYDRA and GOLIAT.
AGA is funded by the UAM. 
Both SA and AGA are further supported by the Ministerio de Ciencia, Innovaci\'{o}n y Universidades (MICIU/FEDER) under research grant PGC2018-094773-B-C32. The authors acknowledge the support of the Spanish Agencia Estatal de Investigacion through the grant “IFT Centro de Excelencia Severo Ochoa by CEX2020-001007-S". IFAE is partially funded by the CERCA program of the Generalitat de Catalunya.

\section*{Data Availability}

The \gpng\ simulations are publicly available on  \url{ftp://honda.ift.uam-csic.es/GOLIAT_PNG_sims/}
For scientific/technical support, please, contact the authors. 
Complementary data requests to the  authors are welcome.



\bibliographystyle{mnras}
\bibliography{FnP_4_PNG} 




\appendix

\section{Further Validation of Fixed PNG simulations}
\label{app:validation}

We discussed in \autoref{sec:ICs} that {\it Fixing} initial conditions modifies the Gaussianity of the initial conditions. This {\it a priori} means that we need to treat with special care {\it Fixed}-PNG simulations. In \autoref{sec:Pk} we already showed that for the halo power spectrum, main focus of this paper and of most interest for future surveys, we do not find any spurious PNG signal due to {\it Fixing} and that {\it Fixed} simulations behave in the same way for $\fnl=100$ as they do for $\fnl=0$. Said otherwise, they give unbiased halo power spectrumn, whereas the error of the {\it Fixed} simulations is reduced. 

The aim of this Appendix is to further validate the usage of {\it Fixed}-PNG by testing other statistics (initial and late bispectra and power spectra of dark matter and halos) and in other configurations (increasing the size or the mass resolution of the simulation).

\subsection{Initial DM power spectrum}

As both PNG and {\it Fixing} modify only the initial conditions, this is a very clean place to look for possible spurious signals. 

\begin{figure*}
    \centering
    \includegraphics[width=0.5\linewidth]{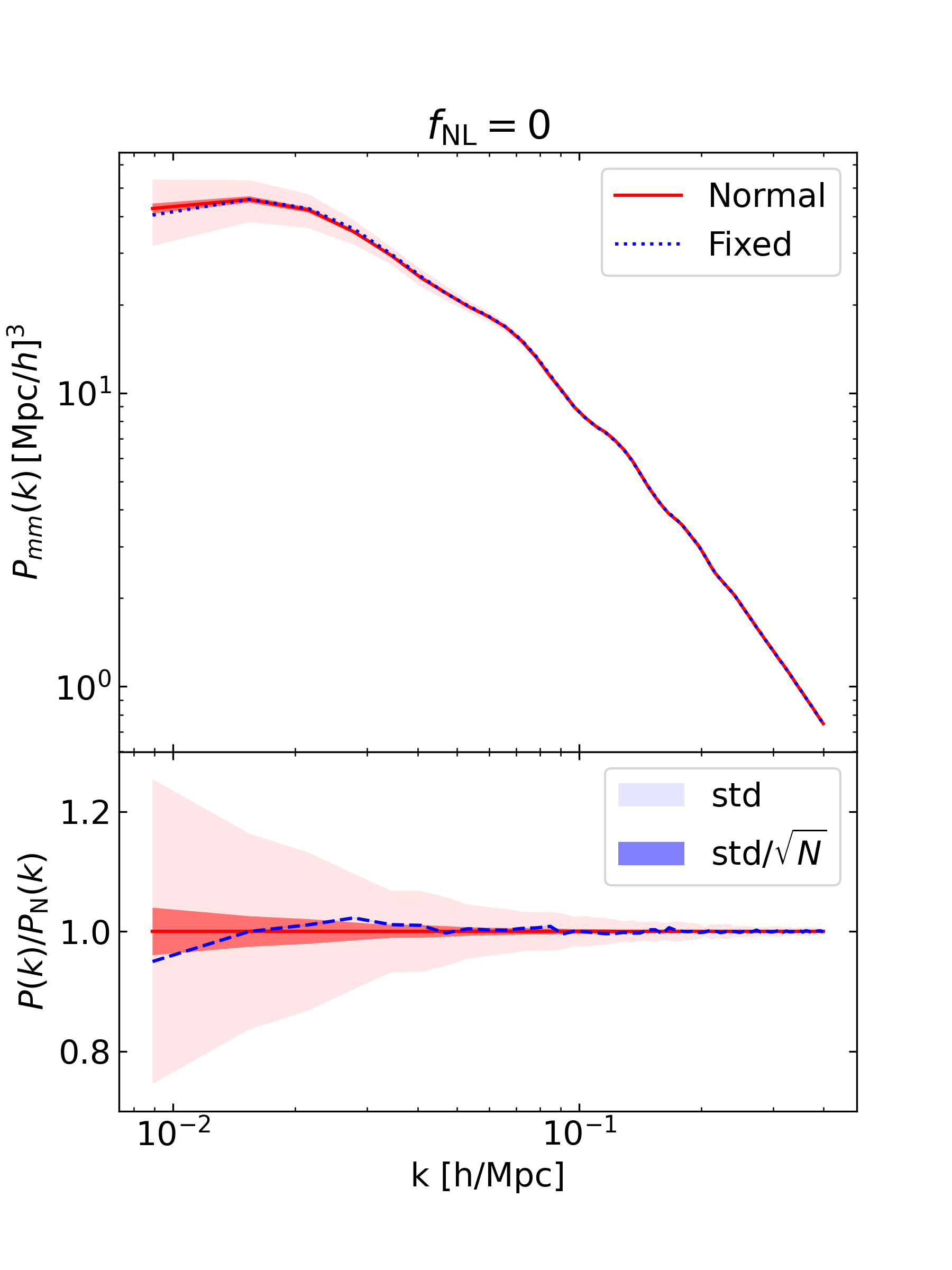}\includegraphics[width=0.5\linewidth]{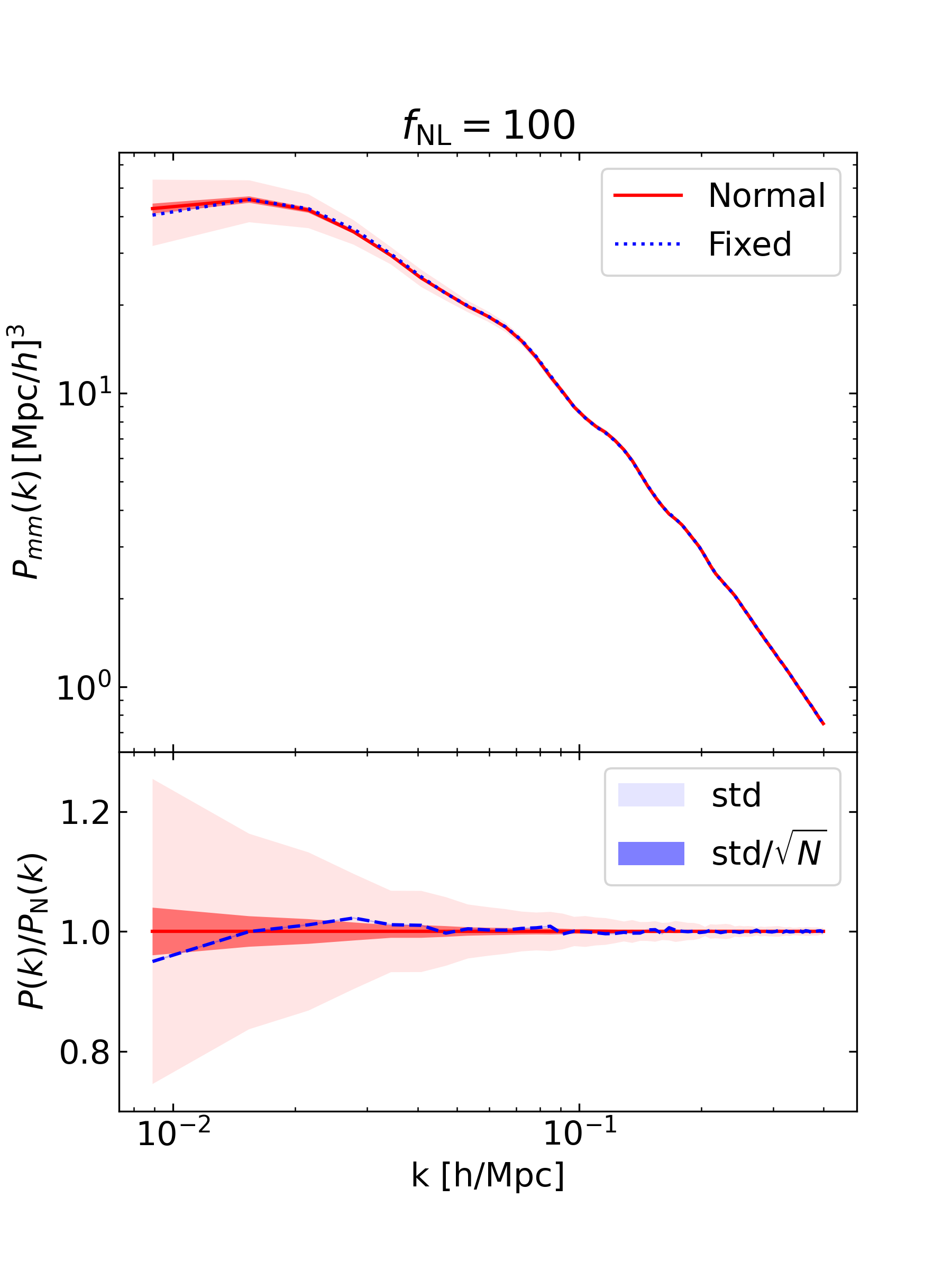}
    \caption{Mean of dark matter power spectra of the initial conditions at $z=32$ for {\it Fixed} and {\it Normal} simulations (Top) and their ratios (Bottom). The standard deviation is shown in a light shaded area, whereas the dark shaded area represents the expected error on the ensamble mean. The left subfigure shows the $\fnl=0$ simulations, whereas the right subfigure shows the $\fnl=100$ case.}
    \label{fig:Pkdm_ICs}
\end{figure*}

In \autoref{fig:Pkdm_ICs}, we show the dark matter power spectra measured in the initial conditions, set at $z=32$. 
We find similar power spectra for $\fnl=0$ and $\fnl=100$, as the PNG is only expected to affect higher order statistics. Also, we see that the variance in the {\it Fixed} power spectra is negligible as expected by construction for this method. 
Looking at the {\it Fixed} to {\it Normal} ratios of dark matter power spectra, $\fnl=100$ simulations appear identical to the $\fnl=0$ case.

\subsection{Initial DM bispectrum}

\begin{figure}
    \centering
    \includegraphics[trim=0 80 25 0, clip
    , width=\linewidth]{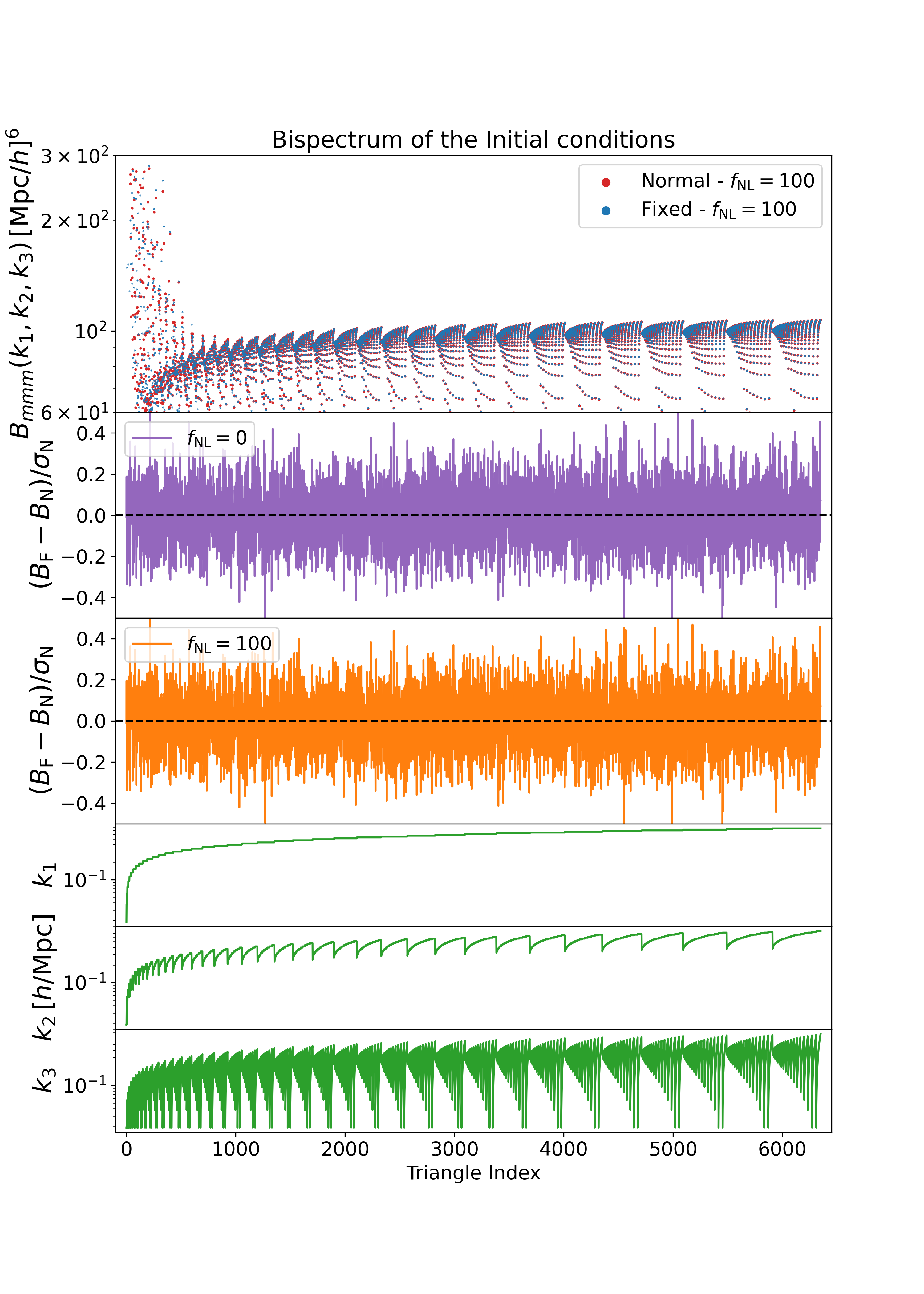}
    \caption{All configurations of the ICs ($z=32$) dark matter bispectrum for $\{k_1,k_2,k_3\}$ being multiples of $3 k_f=0.19 h/{\rm Mpc}$ up to $120 k_f=0.75 h/{\rm Mpc}$, with $k_f$ being the fundamental mode. On the top panel we compare the mean bispectrum of  41 {\it Normal}  simulations to the mean of  41 {\it Fixed} simulations, in both cases with $\fnl=100$. On the next two panels we show the difference between the mean {\it Fixed} bispectrum and the mean {\it Normal} bispectrum over the {\it Normal} standard deviation for the case of $\fnl=0$ and the case of $\fnl=100$, respectively. Finally, the last three panels show the value of $\{k_1,k_2,k_3\}$ for each of the triangles indexed on the $x$-axis. We do not find significant biases and obtain similar results for $\fnl=100$ and $\fnl=0$. 
    }
    \label{fig:Bk_ICs}
\end{figure}

Whereas no signal coming from PNG was expected in the power spectra, we expect to find the net signal induced by $\fnl\neq0$ on the initial dark matter bispectrum.
We compute the bispectrum using the library \textsc{pySpectrum}\footnote{\url{https://github.com/changhoonhahn/pySpectrum}}, considering a binning of $\Delta_k = 3\cdot k_f$ for all the closed triangles up to a $k_{\rm max}$ of $120 k_f$. 
We show all the resulting configurations in \autoref{fig:Bk_ICs}.
Again, we do not find any evidence of spurious PNG signal introduced by the {\it Fixing}. The residuals of the comparison between {\it Normal} and {\it Fixed} simulations look very alike for $\fnl=0$ and $\fnl=100$.

\begin{figure*}
    \centering
    \includegraphics[width=0.5\linewidth]{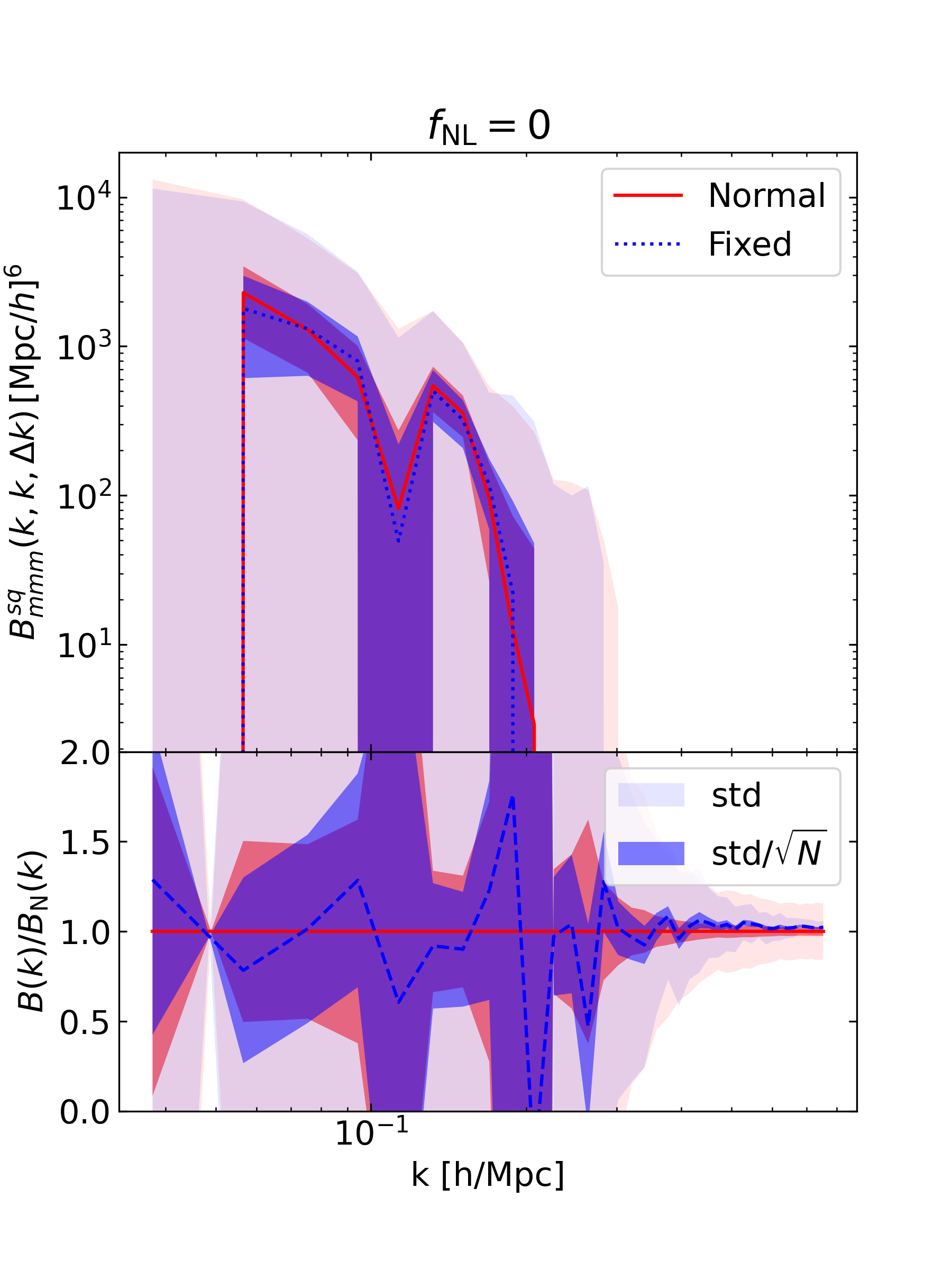}\includegraphics[width=0.5\linewidth]{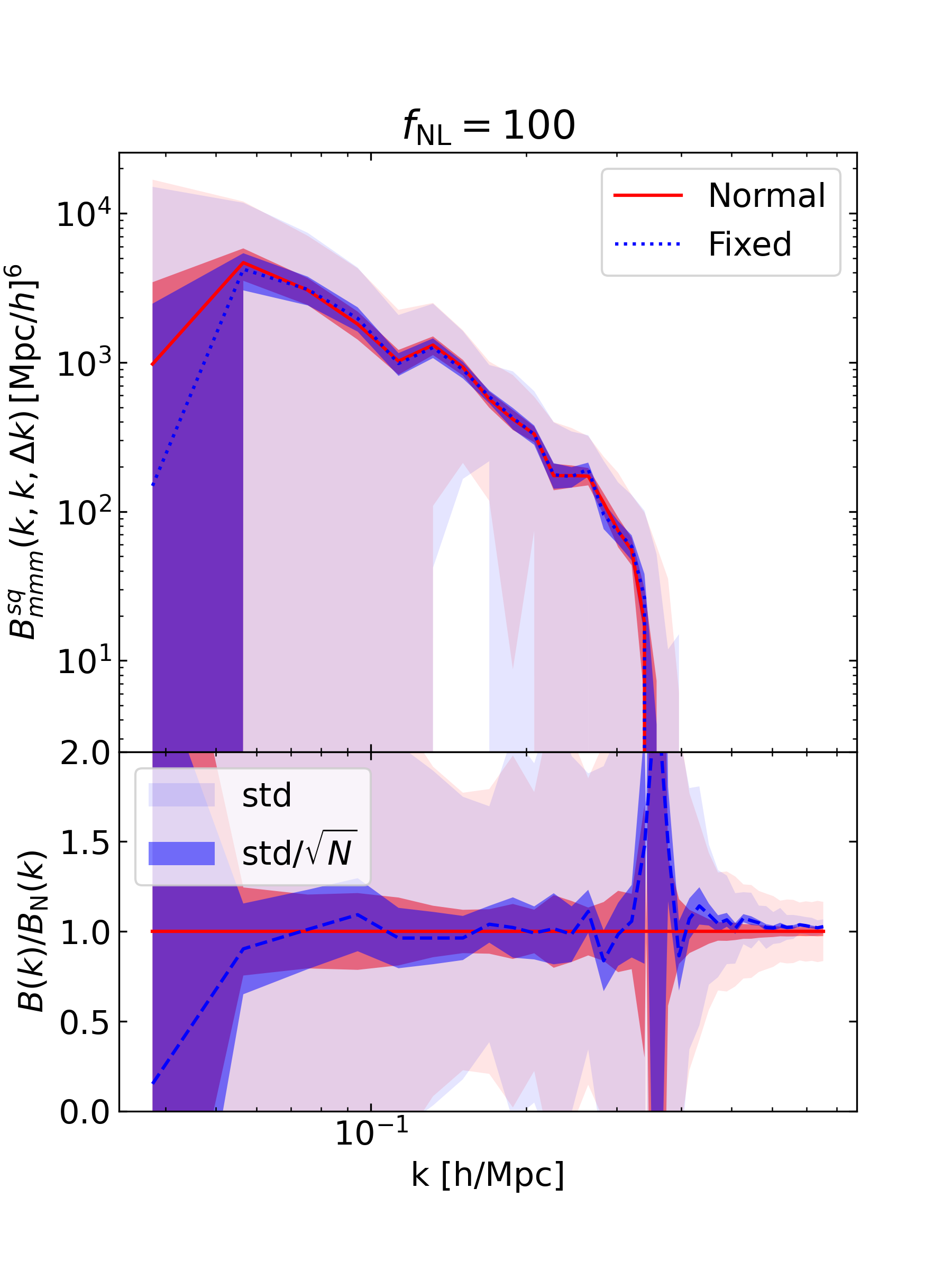}
    \caption{Squeezed bispectrum of the dark matter at the initial conditions ($z=32$) for {\it Fixed} (blue) and {\it Normal} (red) simulations. We show the average (lines) and standard deviations (light shaded) of the 41 realisations studied. On dark-shaded regions we show the estimated error on ensemble average. 
    }
    \label{fig:Bk_sq_ICs}
\end{figure*}

In \autoref{fig:Bk_sq_ICs}, we focus on the squeezed configuration, defined by triangles having a small $k$ and two large $k$. In this case, we fix $k_3=\Delta_k=3k_f$ and $k_1=k_2=k$, leaving $k$ as a variable. This configuration is where the local-PNG is expected to create a signal, not present for the Gaussian case. Indeed, for this configuration, we find a significant bispectrum signal for the ensemble of $\fnl=100$ simulations, whereas, for $\fnl=0$ the signal is comparable with the noise. 
As in previous tests, we do not find differences between the {\it Fixed} and the {\it Normal} simulations. Hence, we do not find any evidence of spurious local-PNG signal. 

\subsection{Late DM power spectrum}

\begin{figure*}
    \centering
    \includegraphics[width=0.5\linewidth]{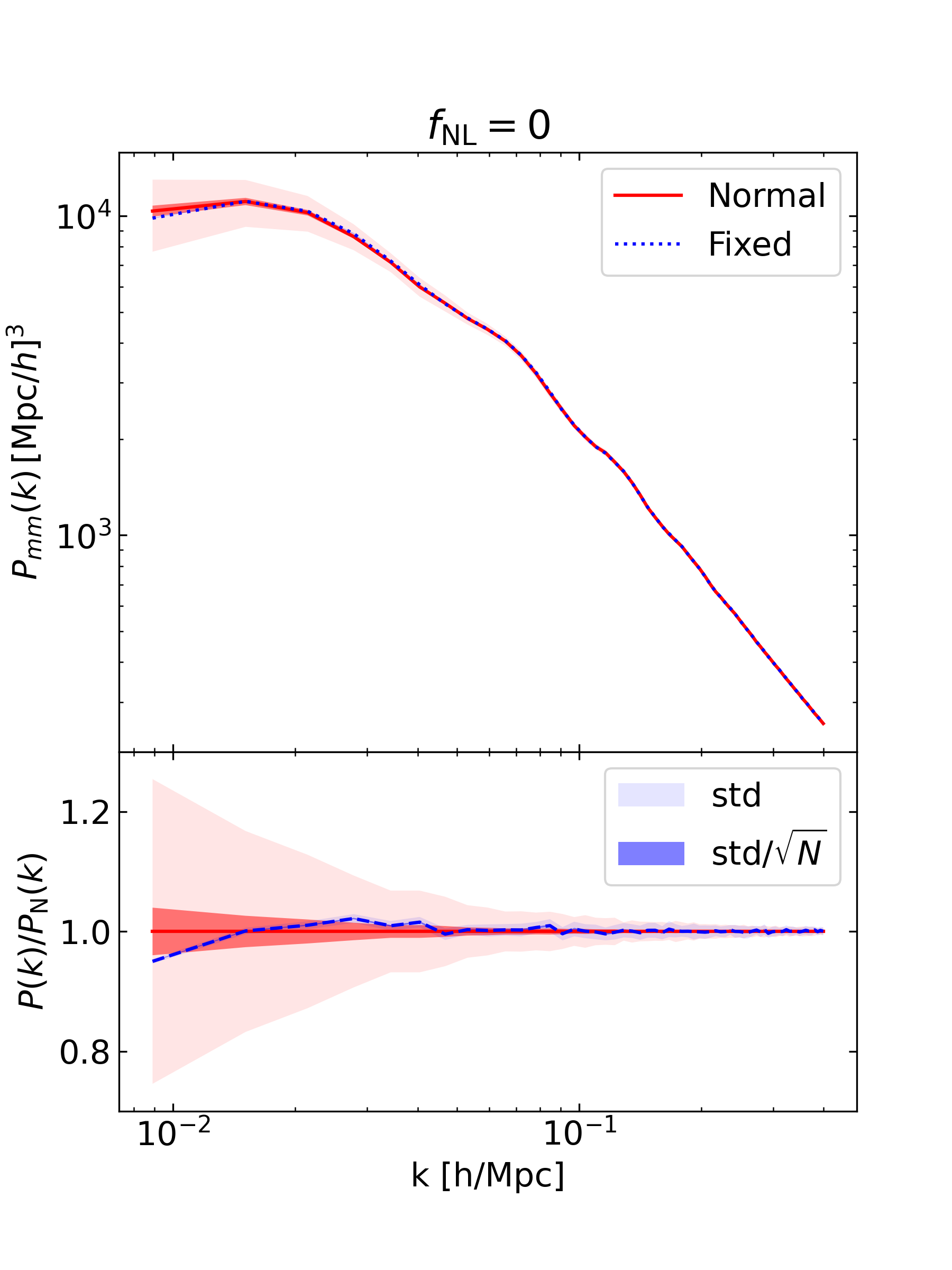}\includegraphics[width=0.5\linewidth]{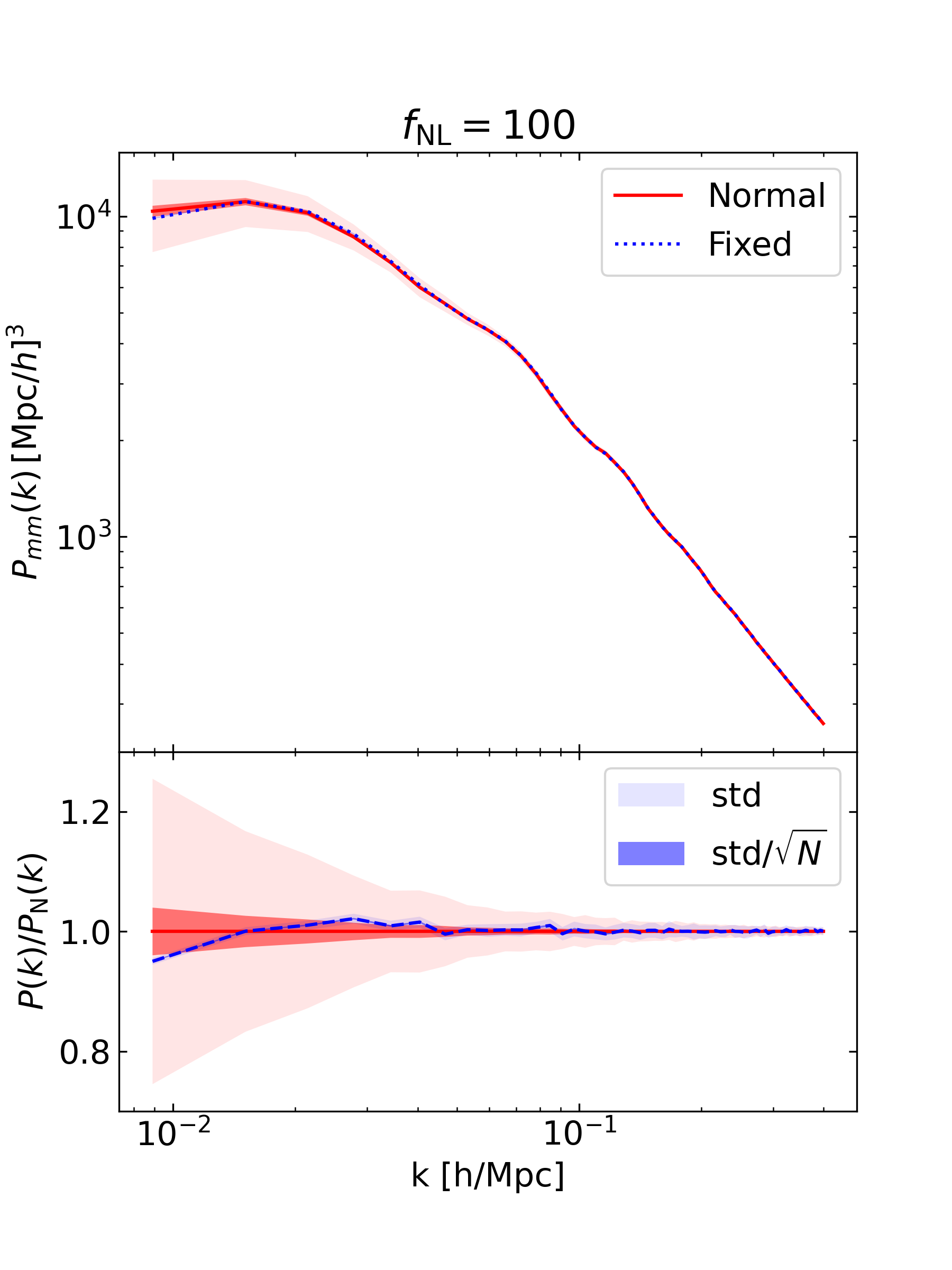}
    \caption{Dark matter power spectrum at redshift $z=1$. We compare the $\fnl=0$ case (left) to the $\fnl=100$ simulations (right), and the {\it Fixed} simulations (blue) to th {\it Normal} ones (red). We find that the {\it Fixed} to {\it Normal} ratios are very similar for the $\fnl=0$ and $\fnl=100$ cases.}
    \label{fig:Pkdm_z1}
\end{figure*}

Whereas we have already studied in detail the initial conditions, where the PNG signal and the {\it Fixing} are input to the simulations, we now check similar statistics at late time, in order to search for possible spurious signals. 
We start by studying in \autoref{fig:Pkdm_z1} the dark matter power spectrum at  $z=1$, redshift of study for the main body of the paper. Again, we find the {\it Fixed} and {\it Normal} simulations compatible and their ratio nearly indistignuishable for the $\fnl=0$ and $\fnl=100$ cases. Thus, we do not find any spurious signal introduced by {\it Fixing} the PNG simulations.

\subsection{Late DM bispectrum}
\begin{figure}
    \centering
    \includegraphics[trim=0 90 35 0, clip, width=\linewidth]{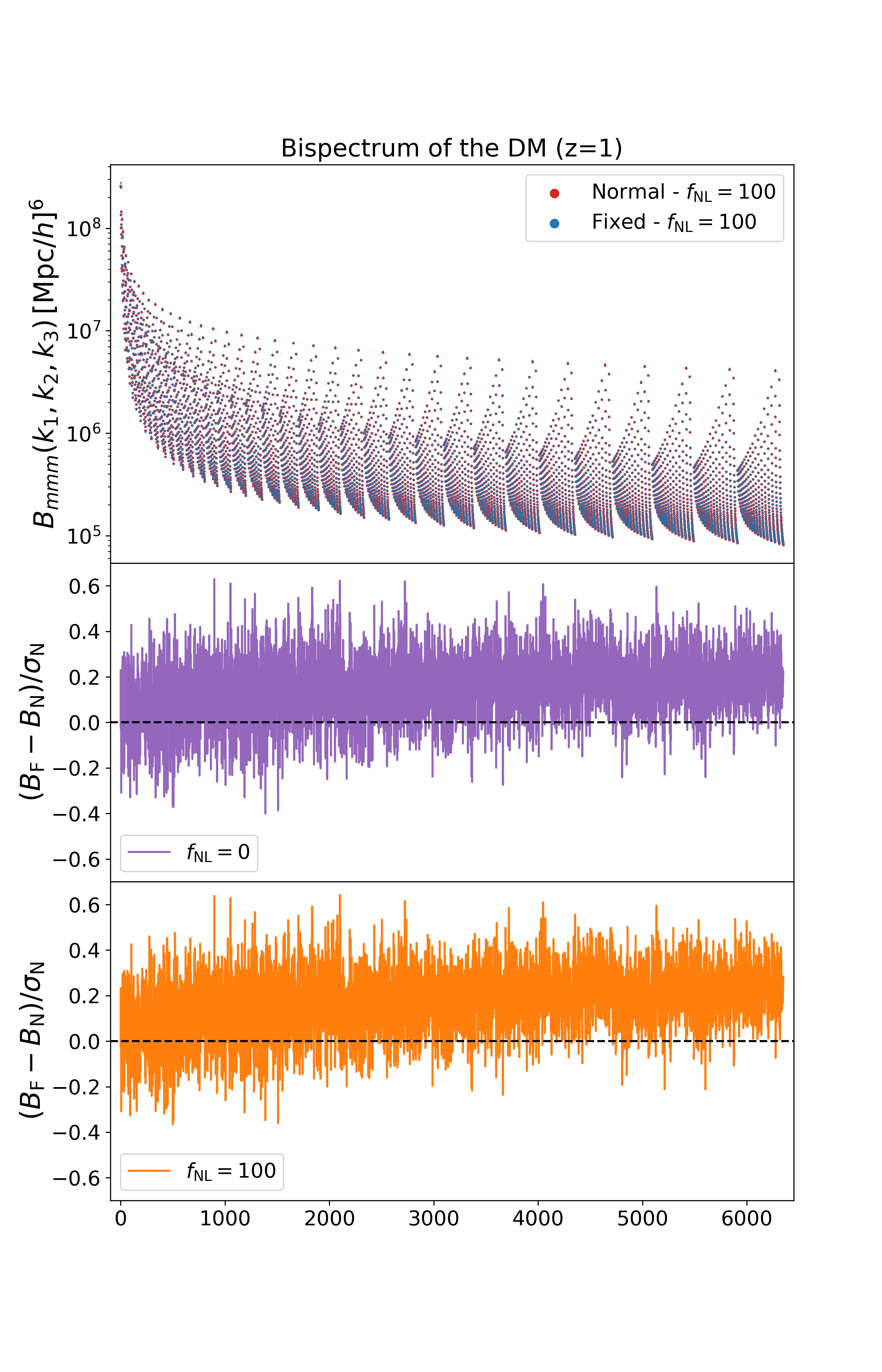}
    \caption{All configurations of the late ($z=1$) dark matter bispectrum for $\{k_1,k_2,k_3\}$ being multiples of $3 k_f=0.19 h/{\rm Mpc}$ up to $120 k_f=0.75 h/{\rm Mpc}$, with $k_f$ being the fundamental mode. On the top panel we compare the mean bispectrum of  41 {\it Normal}  simulations to the mean of  41 {\it Fixed} simulations, in both cases with $\fnl=100$. On the next two panels we show the difference between the mean {\it Fixed} bispectrum and the {\it Normal} bispectrum over the {\it Normal} standard deviation for the case of $\fnl=0$ and the case of $\fnl=100$, respectively. The $x$-axis represents the index of the triangle, each of them with a different $\{k_1,k_2,k_3\}$, as shown in \autoref{fig:Bk_ICs}. We find the $\fnl=0$ and the $\fnl=100$ simulations to behave in the same way when {\it Fixing} the initial conditions.}
    \label{fig:Bk_z1}
\end{figure}

We now put our attention on the dark matter bispectrum at redshift $z=1$. 
First, we remind the reader that this signal is expected to be dominated by {\it late} non-Gaussianities induced by non-linear gravitational evolution of density perturbations. 
Also, that the different triangle configurations will be highly correlated. 
The main conclusion of this subsection can be drawn from the residual plots at the bottom of \autoref{fig:Bk_z1}, where we find the same pattern for $\fnl=0$ and $\fnl=100$. Again, pointing us to the conclusion that the {\it Fixed} initial conditions can be used in the same way for local-PNG as we use them for Gaussian cosmologies.

We do appreciate an offset with respect to a null residual for both $\fnl=0$ and $\fnl=100$. However, this is at the level of $\sim0.2\sigma$ and we find that it is dominated by configurations with $k_i>0.3 h/{\rm Mpc}$. We expect these configurations to be highly correlated one to another and we do not consider these offsets to be significant. Additionally, as the residuals behave similarly for $\fnl=0$ and $\fnl=100$, we do not consider that any of this indicates that the PNG simulations need special treatment. 

For the squeezed configuration (as in \autoref{fig:Bk_sq_ICs}), we found again nearly indistinguishable results for the {\it Fixed} to {\it Normal} ratios for $\fnl=0$ and $\fnl=100$. 
We omit this figure for brevity. 

\subsection{Late halo bispectrum}

\begin{figure}
    \centering
    \includegraphics[trim=0 90 35 0, clip, width=\linewidth]{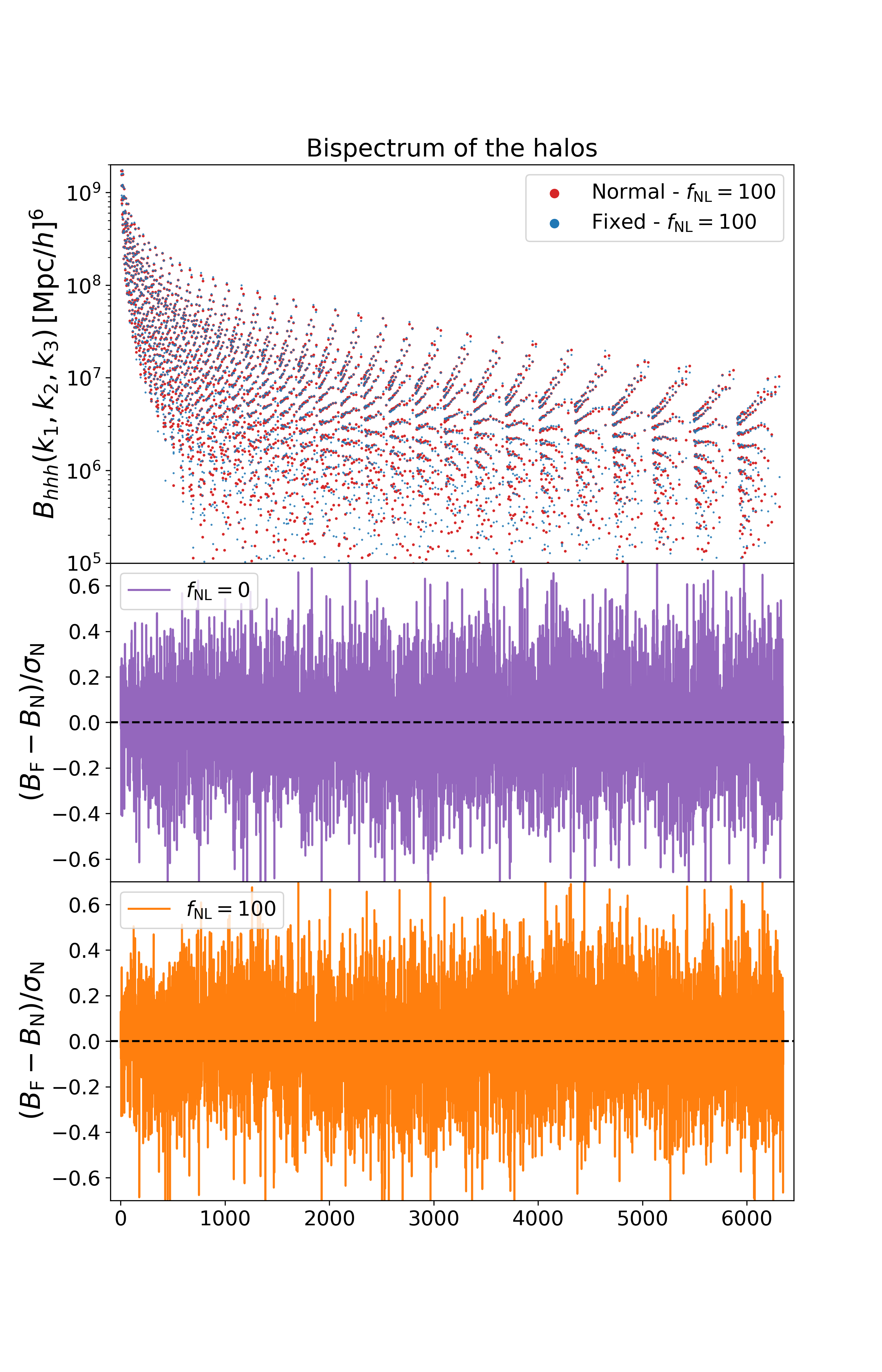}
    \caption{All configurations of the late ($z=1$) halo bispectrum for $\{k_1,k_2,k_3\}$ being multiples of $3 k_f=0.19 h/{\rm Mpc}$ up to $120 k_f=0.75 h/{\rm Mpc}$, with $k_f$ being the fundamental mode. On the top panel we compare the mean bispectrum of  41 {\it Normal}  simulations to the mean of  41 {\it Fixed} simulations, in both cases with $\fnl=100$. On the next two panels we show the difference between the mean {\it Fixed} bispectrum and the {\it Normal} bispectrum over the {\it Normal} standard deviation for the case of $\fnl=0$ and the case of $\fnl=100$, respectively. The $x$-axis represents the index of the triangle, each of them with a different $\{k_1,k_2,k_3\}$, as shown in \autoref{fig:Bk_ICs}.}
    \label{fig:Bkhalos_z1}
\end{figure}

\begin{figure*}
    \centering
    \includegraphics[width=0.5\linewidth]{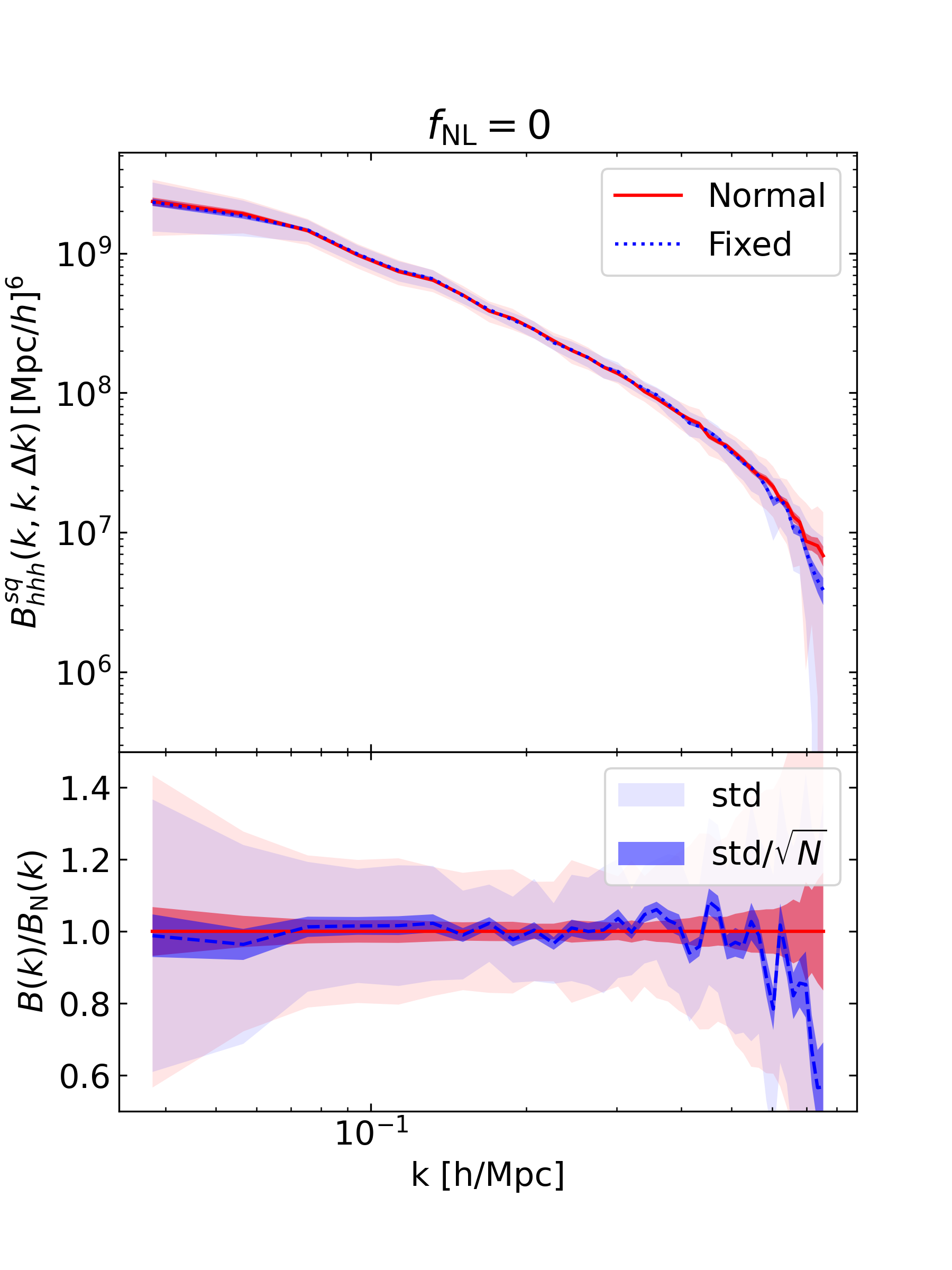}\includegraphics[width=0.5\linewidth]{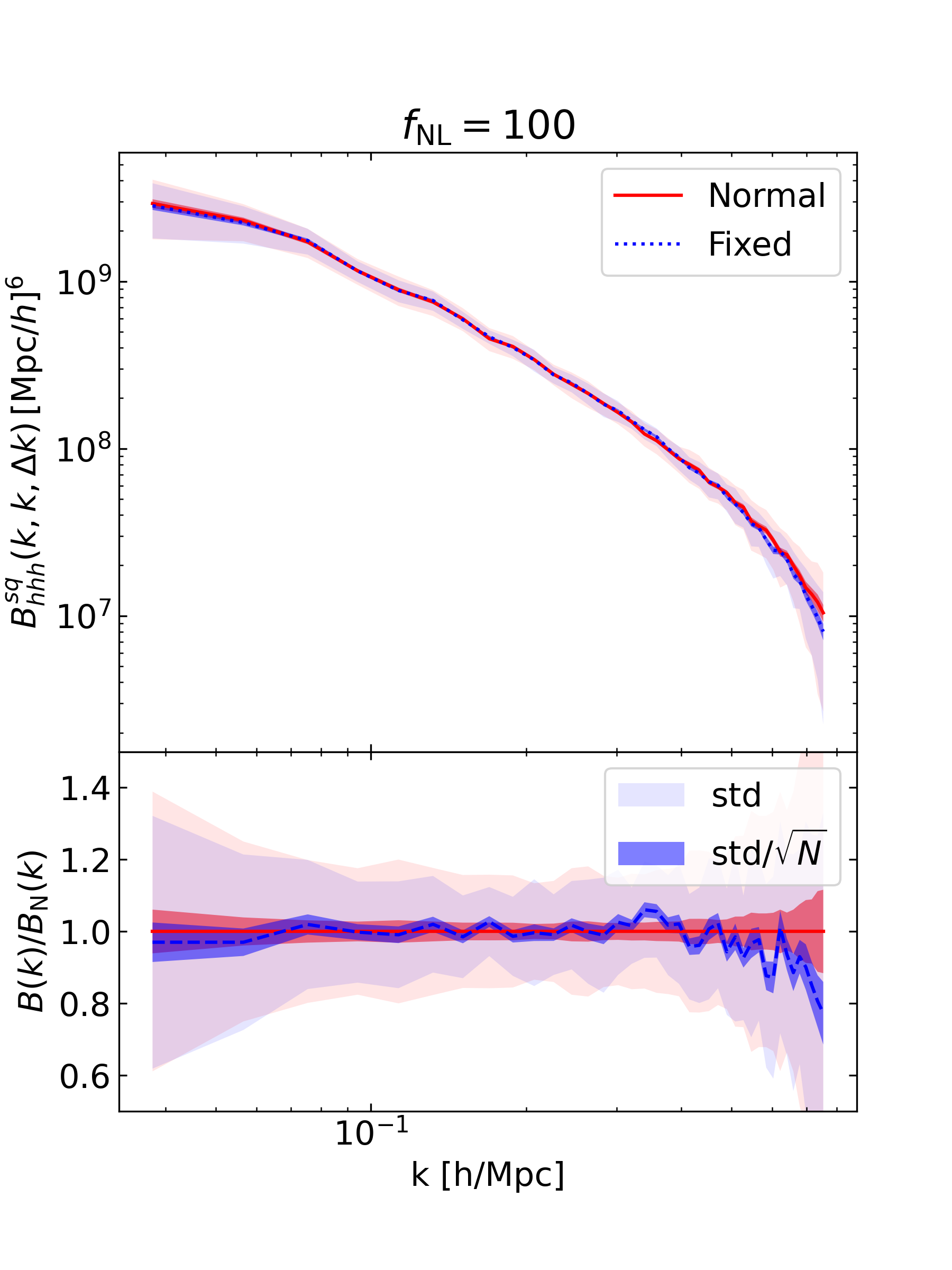}    
    \caption{Squeezed halo bispectrum of {\it Fixed} and {\it Normal} simulations (top), together with their ratio (bottom). We do not find biases induced by {\it Fixing} for either $\fnl=0$ (left) or $\fnl=100$ (right). }
    \label{fig:Bkhalos_sq_z1}
\end{figure*}

We now come back to studying the clustering of halos (at late times, in this case $z=1$). As we validated the halo power spectrum in the main text (\autoref{sec:Pk}) we now focus in another promising observable \citep[see e.g. ][]{Cabass_2022}: the halo bispectrum (as a proxy for galaxy bispectrum).
We show all the configurations in \autoref{fig:Bkhalos_z1} and we focus on the squeezed ones in \autoref{fig:Bkhalos_sq_z1}. In the latter, we can appreciate the signal introduced by local-PNG on this configuration. Besides that, we do not find any evidence of biasing this statistics, and we find the $\fnl=0$ and $\fnl=100$ residuals to behave similarly.

\subsection{Late halo power spectrum: higher resolution or larger box}
\label{sec:highN}

In the main text, we validated the usage of {\it Fixed}-PNG simulations to get unbiased halo power spectra in our default configuration (\autoref{tab:cosmo})
In previous subsections of this Appendix, we have looked at other statistics of the same simulations, not finding any evidence of getting biased statistics by {\it Fixing} local-PNG simulations. 
Finally, here, we study again the halo power spectrum, this time when changing the mass resolution or the box size of our simulations.

\begin{figure*}
    \centering
    \includegraphics[width=0.5\linewidth]{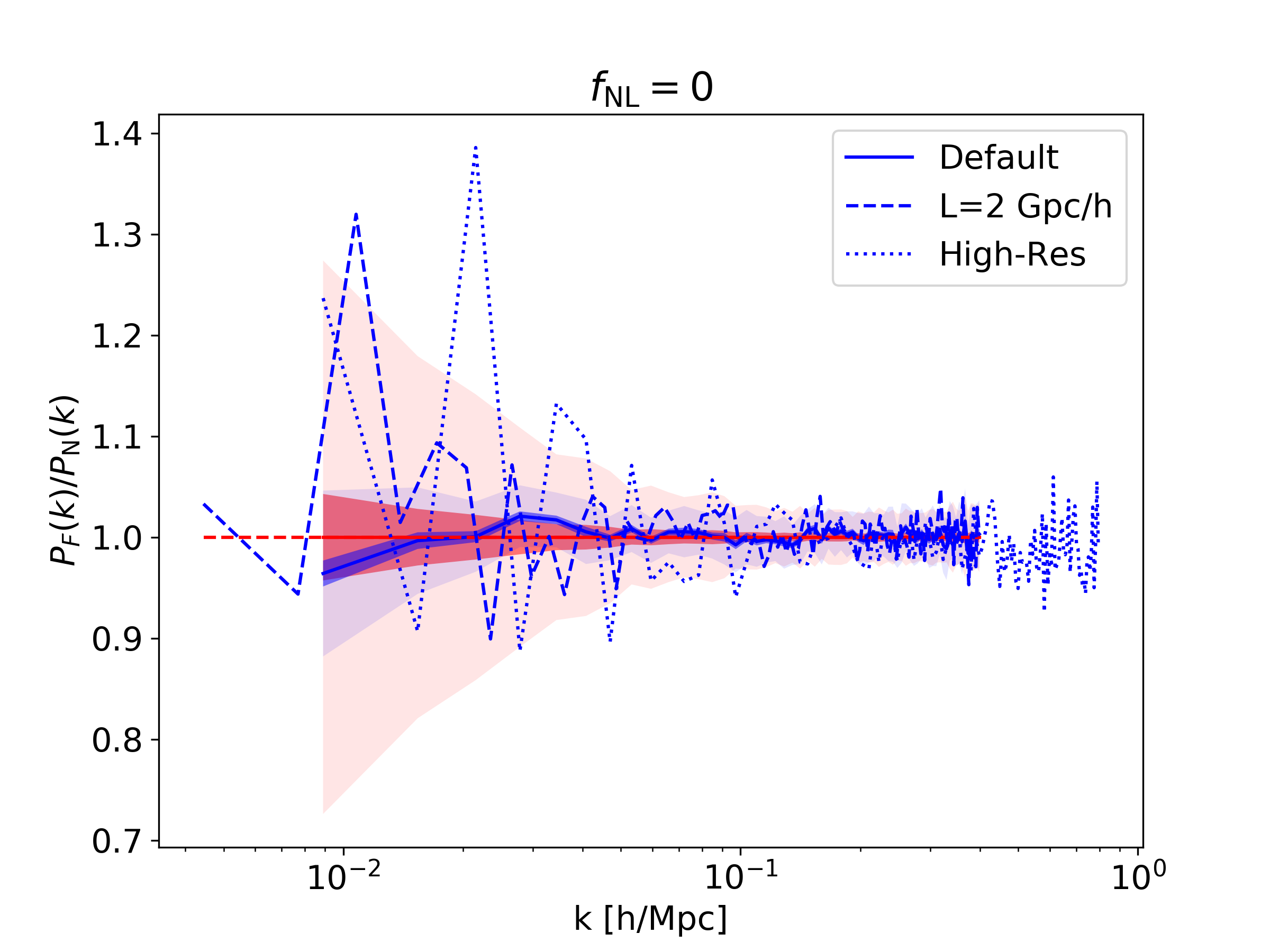}\includegraphics[width=0.5\linewidth]{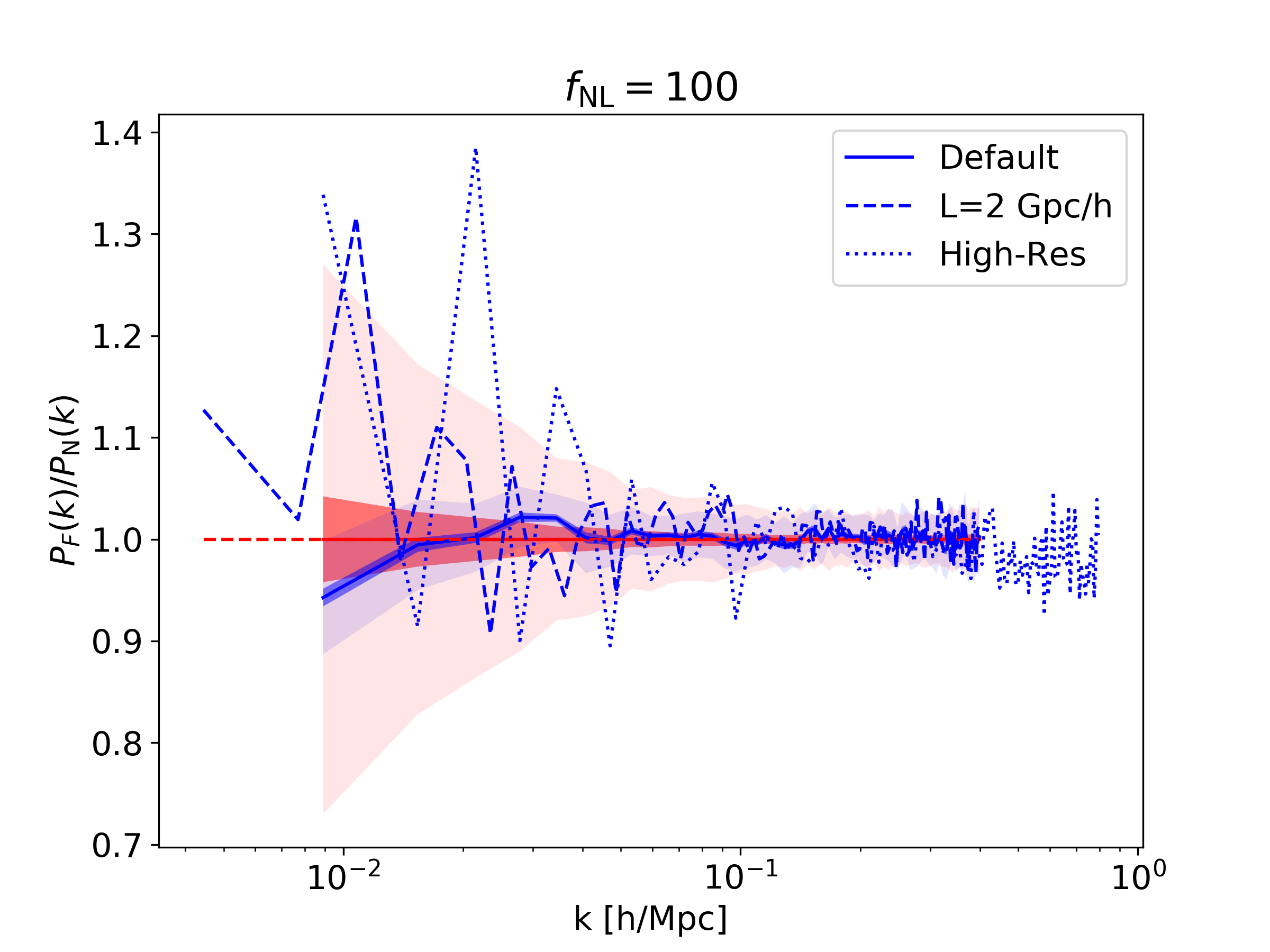}
    \caption{{\it Fixed} to {\it Normal} halo power spectra ratios. On solid lines and light-shaded regions we show the mean and standard deviation of the 41 default ($L=1{\rm Gpc}/h$, $N=512^3$) configuration simulations described in the main text (i. e., they replicate the bottom panels of \autoref{fig:Pk}). The dotted line represents the large box-run ($L=2{\rm Gpc}/h$, $N=1024^3$) and the dashed line represents the  higher resolution run ($L=1{\rm Gpc}/h$, $N=1024^3$). The latter ones only represent one realisation of each type and were only discussed in \autoref{sec:highN}. }
    \label{fig:Pk1024}
\end{figure*}

We ran a few new simulations with 8 times more particles ($N=1024^3$) for the same box configuration (`High-Res') and with double the box size ($L=2{\rm Gpc}/h$, keeping the same mass resolution). In this case, in order to keep the computing time short, we only ran one seed for each of the four combinations with {\it Normal} and {\it Fixed} ICs for both $\fnl=0$ and $\fnl=100$.
In \autoref{fig:Pk1024}, we plot again the {\it Fixed}  to {\it Normal} ratios shown at the bottom of \autoref{fig:Pk} (labeled as `default'), and overplot the new results with higher resolution and larger box. Taking into account that we are only plotting one realisaton for the new cases (hence, with the light-shaded region as reference uncertainty), we find that both cases follow the same trend as our default case and, again, very similar results for $\fnl=100$ and $\fnl=0$.

Remarkably, the largest scales shown for the $L=2{\rm Gpc}/h$ case are already dominated by $\fnl^{\ \ \ 2}$ terms of the halo power spectrum, being able to probe new regimes of the validity of our method. 
Additionally, in a follow-up study \citep{PNGUNITsimXL}, we have further validated these findings (that we still recover unbiased halo power spectra at larger scales and higher mass resolutions) with 10 {\it Fixed} and 100 {\it Normal} \textsc{FastPM} simulations \citep{fastpm} with $L=3{\rm Gpc}/h$ and $N=2560^3$ (we leave the full description of these simulations to a follow-up paper as their  configuration is very different to \autoref{tab:cosmo}).

\bsp	
\label{lastpage}
\end{document}